\newcommand{\PaperTitle}{PixelConfig: Longitudinal Measurement and Reverse-Engineering of Meta Pixel Configurations}
\newcommand{\PaperNumber}{81}

\documentclass[9pt,sigconf,letterpaper,acm]{acmart} % IMC 2025

\AtBeginDocument{%
  }

\copyrightyear{2026}
\acmYear{2026}
\setcopyright{rightsretained}
\acmConference[IMC '26]{Proceedings of the 2026 ACM Internet Measurement Conference}{October 12--16, 2026}{Karlsruhe, Germany}
\acmBooktitle{Proceedings of the 2026 ACM Internet Measurement Conference (IMC '26), October 12--16, 2026, Karlsruhe, Germany}\acmDOI{}
\acmISBN{}
\usepackage{hyperref} % Include hyperref package for hyperlinks
\usepackage{url}
\usepackage{graphicx} % For scaling
\usepackage{array} % For table column width adjustment
\usepackage{booktabs}
\usepackage{multirow}
\usepackage{makecell}
\usepackage{subcaption}
\usepackage{soul}

\usepackage{listings} % For code formatting
\usepackage{xcolor}   % For coloring the code
\usepackage{tcolorbox}% For template quoting
\usepackage{color}
\definecolor{dkgreen}{rgb}{0,0.6,0}
\definecolor{gray}{rgb}{0.5,0.5,0.5}
\definecolor{mauve}{rgb}{0.58,0,0.82}
\definecolor{mygray}{gray}{0.4}
\definecolor{myblue}{rgb}{0.071, 0.125, 0.796}

\definecolor{customgreen}{HTML}{82B366}
\definecolor{customred}{HTML}{EA6B66}
\definecolor{customyellow}{HTML}{D79B00}
\definecolor{customblue}{HTML}{6C8EBF}

\lstdefinelanguage{javascript}{
  keywords={break,case,catch,continue,debugger,default,delete,do,else,false,finally,for,function,if,in,instanceof,new,null,return,switch,this,throw,true,try,typeof,var,void,while,with},
  sensitive=true,
  comment=[l]{//},
  morecomment=[s]{/*}{*/},
  morestring=[b]",
  morestring=[b]',
}

\usepackage{cleveref}
\crefformat{section}{\S#2#1#3}
\crefformat{subsection}{\S#2#1#3}
\crefformat{subsubsection}{\S#2#1#3}

\begin{document}

\thispagestyle{plain}
\pagestyle{plain} % Apply page number style to all pages

\title{\PaperTitle}

%\begin{comment}

\author{Abdullah Ghani}
\affiliation{%
  \institution{Lahore University of Management Sciences, Pakistan}
  \country{}
}
\email{abdullah.ghani@lums.edu.pk}

\author{Yash Vekaria}
\affiliation{%
  \institution{University of California, Davis}
  \country{}
}
\email{yvekaria@ucdavis.edu}

\author{Zubair Shafiq}
\affiliation{%
  \institution{University of California, Davis}
  \country{}
}
\email{zubair@ucdavis.edu}

\renewcommand{\shortauthors}{Ghani et al.}

\begin{abstract}
Tracking pixels are used to optimize online ad campaigns through personalization, re-targeting, and conversion tracking. 
Past research has primarily focused on detecting the prevalence of tracking pixels on the web, with limited attention to how they are configured across websites.
A tracking pixel may be configured differently on different websites.
In this paper, we present a differential analysis framework: \texttt{PixelConfig}, to reverse-engineer the configurations of Meta Pixel deployments across the web.
Using this framework, we investigate three types of Meta Pixel configurations: activity tracking (i.e., what a user is doing on a website), identity tracking (i.e., who a user is or who the device is associated with), and tracking restrictions (i.e., mechanisms to limit the sharing of potentially sensitive information).

Using data from the Internet Archive's Wayback Machine, we analyze and compare Meta Pixel configurations on 18K health-related websites with a control group of the top 10K websites from 2017 to 2024.
We find that activity tracking features, such as automatic events that collect button clicks and page metadata, and identity tracking features, such as first-party cookies that are unaffected by third-party cookie blocking, reached adoption rates of up to 98.4\%, largely driven by the Pixel's default settings.
We also find that the Pixel is being used to track potentially sensitive information, such as user interactions related to booking medical appointments and button clicks associated with specific medical conditions (e.g., erectile dysfunction) on health-related websites.
Tracking restriction features, such as  Core Setup, are configured on up to 34.3\% of health websites and 8.7\% of control websites.
However, even when enabled, these tracking restriction features provide limited protection and can be circumvented in practice. 
\end{abstract}



% \received{20 November 2025}
% \received[revised]{4 February 2026}
% \received[accepted]{2 March 2026}

\maketitle

\section{Introduction}
Online advertising platforms such as Google, Meta, and TikTok provide tracking pixels that advertisers can install on their websites. 
The information collected through these tracking pixels is used to   personalized ad campaigns, re-target users \cite{Meta2025Retargeting}, reach lookalike users \cite{Meta2025LookalikeAudience}, or  perform conversion optimization \cite{Google2025Pmax, Meta2025Advantageplus}.
Today, almost every business has an online presence and most of them run ad campaigns to reach new customers and engage with the existing ones. 
This means that businesses install tracking pixels provided by online advertising platforms on their websites to track their customers \cite{Lerner2016InternetjonesUSENIX}. 
In fact, over 90\% of websites include at least one tracking pixel today \cite{Dambra2022trackingUSENIX,Englehardt2016TrackingCCS}.

Tracking pixels have advanced in their capabilities over the years. 
Early tracking pixels were simple 1×1 image elements.
When a user visited a webpage where a  tracking pixel was installed, the user's IP address, cookies, and information about the page's URL was shared with the tracker. 
Modern tracking pixels (now often referred to as tags instead of pixels) increasingly rely on JavaScript to---either automatically or through additional configuration---collect a richer set of information made available via various web APIs.
They can be customized in a variety of ways to use different tracking features.
For example, modern tracking pixels can be configured to use first-party cookies \cite{Meta2025FPCookiesDocumentation, Tiktok2025FPCookiesDocumentation} or share information about specific user activities such as button clicks on a page \cite{Google2025ClicksDocumentation, Meta2025ClicksDocumentation}.

The research community has extensively studied the prevalence of tracking pixels \cite{vekaria2025sok,Narayanan2017WebtapSpringer,Englehardt2016TrackingCCS,Lerner2016InternetjonesUSENIX, Bekos2023HitchikerWebconf}.
However, these prior works have primarily focused on detecting the presence of tracking pixels across the web, overlooking the fact that the same tracking pixel installed on two websites may be configured differently. 
Since modern JavaScript-based tracking pixels offer a myriad of configurable features, simply detecting their presence on a website is insufficient to understand the full scope of their tracking capabilities.

To address this gap, we aim to study different configurations of the same tracking pixel across the web.
However, this is technically challenging. 
First, public documentation of tracking pixels is often vague and incomplete.
Second, studying tracking pixel configurations within the advertiser portal could be helpful in theory; however, infeasible in practice as it requires access to the advertiser accounts.
Third, dynamic network traffic analysis of tracking pixels through website crawling suffers from completeness issues because it requires running a deep crawler to identify relevant pages on a website and simulating robust user interactions such as filling form fields. 
Additionally, dynamic analysis restricts the scope to live installation of pixel on the website, limiting retrospective analysis of configuration changes over time.
Fourth, static analysis of JavaScript source code to study tracking pixel configuration is challenging since the code is typically obfuscated and minified.

To address these challenges, we present \texttt{PixelConfig}---a reverse-engineering framework that combines static and dynamic approaches to perform a differential analysis of tracking pixel configurations deployed across the web and over time.
This involves iteratively patching different tracking pixel configurations of a given website for ablation analysis of specific parts of the code pertaining to specific configurations.
Next, we compare network traffic of the original pixel configuration code and the patched one replayed in the browser. 
We further create a developer account and install the tracking pixel on a test website to enumerate different configurations.
These differences in the pixel's source code and network traffic allow us to pinpoint the exact part of the pixel source code that reflects a specific pixel configuration.

\begin{figure*}[t]
    % \vspace{-1mm}
    \centering
    \includegraphics[width=1\linewidth]{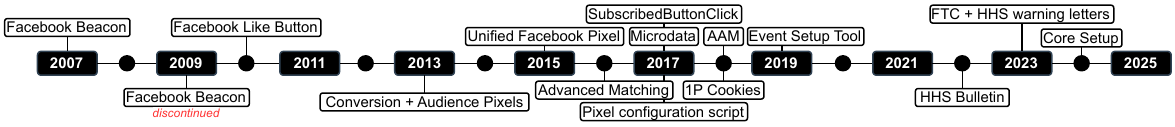}
    % \vspace{-5mm}
    \caption{Timeline depicting key events in the evolution of Meta Pixel: features, configurations, regulatory actions.} 
    % \textcolor{customgreen}{Green} represents introduction of a feature; \textcolor{customred}{Red} represents shutdown of a feature; \textcolor{customblue}{Blue} represents introduction of pixel configuration; and \textcolor{customyellow}{Yellow} represents media and regulatory actions.}
    \label{fig:meta-pixel-timeline}
    \vspace{-2mm}
\end{figure*}

In this paper, we apply \texttt{PixelConfig} to investigate configuration of Meta Pixel (from Meta, formerly known as Facebook) across the web from 2017 to 2024.
Building on the approach introduced by Lerner et al. \cite{Lerner2016InternetjonesUSENIX} and most recently used by Bahrami et al. \cite{bahrami2021fingerprintingPETS}, we rely on Meta Pixel's source code installed across websites over the years as archived by the Internet Archive's Wayback Machine \cite{archive2025Waybackmachine}. 
Using the data from the Wayback Machine, we conduct a longitudinal analysis of the configurations of Meta Pixel installed on 18K health websites and a control group of top-10K websites. 
Tracking pixels installed on health websites can collect potentially sensitive information when users search for specific medical conditions, schedule medical appointments, or access medical records. 
Health information is considered highly sensitive \cite{ur2012smart,leon2013matters,melicher2016not} and laws/regulations (e.g., HIPAA \cite{Hipaa1996}) provide heightened protections for health information. 
Using \texttt{PixelConfig}, we study the adoption and configuration of the following three categories of Meta Pixel features:

\noindent $\bullet$ {\textbf{Activity Tracking:}} 
    Meta Pixel tracks user activity---what a user is doing on a website---through about 20 default and standard events such as \texttt{PageView}, \texttt{Purchase}, and \texttt{AddToCart} that can be automatically setup. 
    Meta Pixel also allows advertisers to define custom events.
    These events can be configured in several ways, allowing Meta Pixel to track a wide range of information such as page URL, referrer URL, page title, button text, form fields, etc.

\noindent $\bullet$ {\textbf{Identity Tracking:}} 
    Meta Pixel tracks user identifying information---who a user is or who the device is associated with---using PII (e.g., {email}, {phone}, {firstname}, {lastname}) harvested from web forms, first- and third-party cookies containing account and device identifiers, as well as IP address and user-agent. 
    
\noindent $\bullet$ {\textbf{Tracking Restrictions:}} Meta Pixel has recently introduced tracking restrictions that can be configured to limit the information shared with Meta. 
    These restrictions include recently introduced Core Setup \cite{Meta2025CoreSetup} as well as {blacklisted} and {sensitive} keys to limit tracking by Meta Pixel.

We aim to address the following research questions: 

\noindent {\textbf{\textit{(1) How has Meta Pixel configuration evolved over time?}}}
Over the years, Meta Pixel has offered various new features and capabilities, but how these are adopted and configured remains unknown.
Understanding tracking pixel configurations over the time can shed light on how advertisers have responded to new features introduced by Meta, as well as the impact of new privacy laws, regulations, and enforcement actions.

\noindent {\textbf{\textit{(2) How does 
    Meta Pixel configuration compare across health-related and a control group of top websites?}}} 
    A tracking pixel installed on a generic website (e.g., a news site) presents different privacy risks as compared to the same pixel installed on a health website, due to the sensitive nature of health information and its heightened regulatory protections. % under laws/regulations. 
    The collection and sharing of potentially sensitive health information by tracking pixels has invited scrutiny from regulators such as HHS and FTC~\cite{FTC2023HeartTracking, alder2023warninghipaa}.

Our findings show that the adoption of tracking features of Meta Pixel is driven by its default settings. 
%
% For example, Meta Pixel's automatic event tracking feature that collects button clicks and page metadata information was adopted by up to 98.4\% of the websites, while its first-party cookie feature, which is unaffected by third-party cookie blocking, was also adopted by $98.4\%$ of the websites. 
%
For example, Meta Pixel’s automatic event tracking feature—collecting button clicks and page metadata—was adopted by up to 98.4\% of the websites; similarly, its first-party cookie feature, which remains unaffected by third-party cookie blocking, saw adoption at the same rate.
This is because both of these features were configured by default. 
Adoption of other features such as Automatic Advanced Matching---not turned on by default---was driven by Meta's messaging and nudging. 
%
% For example, Automatic Advanced Matching---involving extraction of identifying information from web form fields---was adopted by nearly 50\% of the websites due to Meta's messaging around its ability to improve the effectiveness of ad campaigns in light of anti-tracking features introduced in web browsers \cite{Meta2025AutoAdv,archive2022AAM}. 
%
Meta Pixel has also introduced  tracking restriction features. 
These include blacklisted and sensitive keys (to block tracking of certain URL parameters) in 2020 and 2021 respectively, and more recently Core Setup in 2023. 
We show that these restrictions are typically enabled or configured by Meta and are adopted more on health websites as compared to control, though overall adoption even on health websites remains low ($25.4\%$ for sensitive keys and $34.3\%$ for Core Setup).
We also find that these restrictions, even when implemented, were not always effective (e.g., not all potentially sensitive URL parameters were blocked and not all Pixels on a health website were placed in Core Setup).

%

% \vspace{-3mm}
\section{Background \& Related Work}
\label{sec:background}

\subsection{Introduction to Meta Pixel}
\label{sec:what-is-meta-pixel}

Meta Pixel (formerly, Facebook Pixel) is a JavaScript-based tracking pixel that allows Meta to track information about certain actions taken by a user visiting a website~\cite{meta2019aboutpixel}. 
The information tracked by Meta Pixel on a website is used to optimize ad campaigns on Meta. 
For example, an advertiser may run ad campaigns to retarget users on Instagram who previously added an item to cart but did not complete the purchase on the website. 
Meta Pixel is currently used on millions of websites \cite{Builtwith2025Metapixel}, accounting for one-quarter of the web according to various measurements \cite{Dambra2022trackingUSENIX,Vekaria2024ThirdpartiesAlmanac,Munir2023CookieCCS,Bekos2023HitchikerWebconf}.

Meta Pixel builds on Meta's earlier tracking tools, which date back more than 15 years as depicted in Figure~\ref{fig:meta-pixel-timeline}.
Beacon was the first tracking tool introduced in 2007 that enabled Facebook to track user activities on non-Facebook websites \cite{dave2007beacon,schiffman2007beaconWired}. 
When a user visited a website where Beacon was installed, the user's activity on the website was (without explicit user interaction) shared on the user's Facebook News Feed. 
It was discontinued in 2009 due to privacy concerns \cite{cbc2009beacon}. 
In 2010, Facebook introduced the Like button that could be installed on non-Facebook websites \cite{roosendaal2010LikebuttonSSRN}. 
Both Beacon and the Like button automatically tracked user activity (e.g., URL of the page and a third-party cookie) but the Like Button required a user to explicitly click on the button for user's website activity to be shared on the user's Facebook News Feed \cite{roesner2012detectingNDSS,roosendaal2010LikebuttonSSRN}.

In 2013, Facebook introduced two advertising-focused pixels that were ultimately merged into a unified Facebook/Meta Pixel \cite{calero2015pixelupgradeJLD,facebook2015pixelWayback,martech2013conversiontracking,meta2015pixelupdateDocs}. 
The Custom Audience Pixel enabled creation of an audience of users on the website that visited a particular URL \cite{Kierbow2019customaudiences}.
The Conversion Tracking Pixel enabled tracking information about specific actions users took on the website via five standard events (i.e., Checkouts, Registrations, Leads, Key Page Views, Adds to Cart) \cite{Loomer2025creatingpixelJLD}.
In 2015, Facebook unified the two pixels with the launch of the Facebook Pixel \cite{facebook2015pixelWayback,calero2015pixelupgradeJLD}.
The unified Facebook Pixel enabled tracking of nine standard events (e.g., \texttt{InitiateCheckout}, \texttt{Purchase}) and define custom events \cite{meta2015pixelupdateDocs,calero2015pixelupgradeJLD}. 
Just like Beacon and the Like Button, Facebook Pixel automatically tracked user activity on non-Facebook websites but with more detailed information and 
% the main purpose being 
to optimize Facebook ad campaigns.

In an update in 2017, Facebook Pixel introduced two new events to automatically collect button clicks (\texttt{SubscribedButtonClick}) and page metadata (\texttt{Microdata}) on non-Facebook websites \cite{archive2017pixel}. 
Unlike previous iterations of the Facebook Pixel, the collection of these automatically collected events did not require any manual configuration from website developers \cite{Pixelyoursite2017Autoevents, Meta2025ClicksDocumentation}. 
Using automatic events, Facebook Pixel now could automatically infer that these events are associated with standard events such as \texttt{Purchase} or \texttt{AddToCart}. 
Beginning with this update, as discussed later in Section~\ref{sec:meta-pixel-configurations}, Facebook Pixel's source code was split into two scripts: a generic \textit{fbevents.js} script \cite{archive2017fbevents} and a \textit{signals/config} script \cite{archive2017configfile} that contains configuration information about a Meta Pixel.

%%
% \vspace{-3mm}
\subsection{How does Meta Pixel track users?}
\label{sec:how-pixel-tracks-users}
At its core, Meta Pixel tracks a user's activity on non-Facebook websites and matches it with the user's Facebook account.

\vspace{0.05in} \noindent $\bullet$ \textbf{Activity Tracking:}
Meta Pixel by default supports the \texttt{PageView} event for each page load. 
In addition, Meta Pixel supports two automatic events---\texttt{SubscribedButtonClick} and \texttt{Microdata} \cite{Pixelyoursite2017Autoevents}. 
The \texttt{SubscribedButtonClick} event captures button click information such as button text. 
The \texttt{Microdata} event captures metadata of the webpage that is defined on the page using OpenGraph, Schema.org, or JSON-LD format. 
Beyond these, Meta Pixel supports 17 predefined standard events, representing common user actions indicative of conversions, such as product searches, views, and purchases. 
Each standard event may include parameters such as product name, product identifier, product category, product price, quantity purchased, etc. 
Meta Pixel also supports custom events, which allow defining and capturing any user action not captured by the standard events \cite{meta2025Pixeladvanceddoc,meta2025Eventsetuptool,meta2025Autoevents}.

Each Meta Pixel event is sent as an HTTP GET/POST request to Meta's \textit{\url{https://www.facebook.com/tr/?id=[PixelID]}} server, where PixelID is the Meta-assigned unique identifier for the pixel instance. 
The payload of each HTTP request includes the page URL in the \texttt{dl} parameter and referrer page URL in the \texttt{rl} parameter \cite{meta2019aboutpixel}. 
Beyond standard URL parameters, event-specific and contextual details are included in \texttt{cd[<parameterKey>]} parameters, where \texttt{parameterKey} may be \texttt{buttonText}, \texttt{buttonFeatures}, \texttt{pageFeat- ures}, and \texttt{formFeatures} for the \texttt{SubscribedButtonClick} event and DataLayer such as OpenGraph, Schema.org, and JSON-LD for the \texttt{Microdata} event. 
Other standard and custom events may also include relevant information in the \texttt{cd[<parameterKey>]} parameters. Lastly, to facilitate user-matching, user data manually sent by advertisers is included in the \texttt{ud[<parameterKey>]}, or the \texttt{udff[<parameterKey>]} if captured automatically through form fields via Automatic Advanced Matching, where common \texttt{parameter Key} values include identifiers such as \texttt{em} (for email) as discussed below in identity tracking.
% , \texttt{fn} (first name), or \texttt{ph} (phone number).

\begin{figure*}[t]
    \centering
    \includegraphics[width=\linewidth]{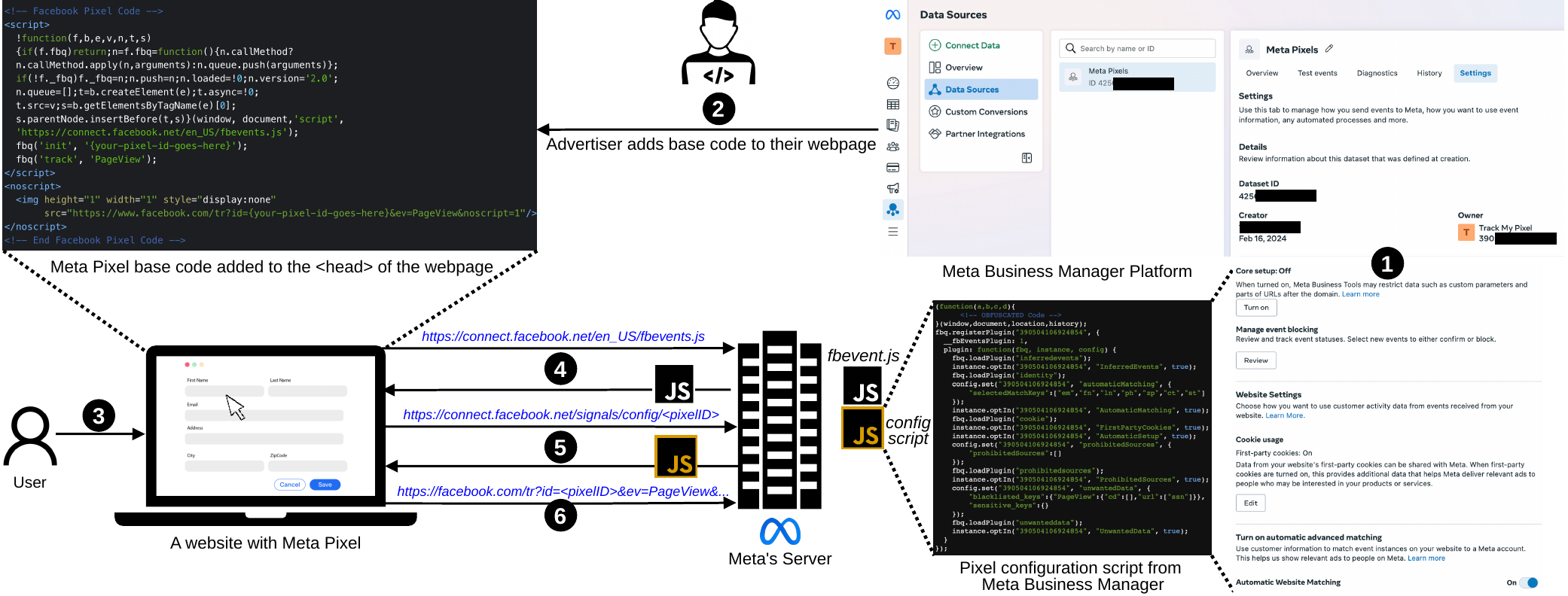}
    % \vspace{-3mm}
    \caption{Steps depicting the process of configuring, loading, and tracking using a Meta Pixel.}
    % \caption{Steps depicting the process of loading a configured Meta Pixel.}
    % \vspace{-2mm}
    \label{fig:loading-meta-pixel}
\end{figure*}

% Command to Compress:
% gs -sDEVICE=pdfwrite -dCompatibilityLevel=1.4 -dPDFSETTINGS=/prepress \
%    -dNOPAUSE -dQUIET -dBATCH -sOutputFile=optimized-configuring-meta-pixel.pdf configuring-meta-pixel.pdf 

%%
\vspace{0.05in} \noindent $\bullet$ \textbf{Identity Tracking:}
Meta Pixel collects three categories of identifying information for each event and matches them to a user's Facebook account using a proprietary matching algorithm \cite{Veerabathini2024AutoadvCustomerlabs,Meta2025AutoAdv}. 
\textit{First}, Meta Pixel automatically collects third-party cookies set on the \textit{facebook.com} domain with each event's HTTP request. 
These include the \texttt{c\_user} cookie that stores the Facebook user ID in the plaintext, the \texttt{fr} cookie that contains encrypted Facebook user ID and a browser ID, and the \texttt{datr} cookie that contains a browser ID \cite{davis2011FacebookIrelandAudit, facebookCookiesPolicy}.
Given some browsers restrict third-party cookies \cite{cj2019ITP,Camp2019ETPMozilla}, Meta Pixel also stores first-party cookies (i.e., set on the advertiser's domain), \texttt{\_fbp} and \texttt{\_fbc}, since 2018~\cite{Digiday2018FPcookies}. 
The \texttt{\_fbp} cookie enables \textit{same-site} tracking while the \texttt{\_fbc} cookie enables cross-site tracking by storing the \texttt{fbclid} click ID that is added as a URL parameter by Facebook during any Facebook to non-Facebook navigation \cite{Bekos2023HitchikerWebconf}. 
The third-party cookies are shared with Facebook in the Cookie header while the first-party cookies are shared in parameters. 
\textit{Second}, in addition to cookies, Meta Pixel automatically collects IP address, user agent, and other device properties such as screen width and height. 
While account and device identifiers in cookies are deterministically matched to a Facebook user, IP address, user agent, and other device properties are probabilistically matched \cite{sen2023IdentitySolutions}.
\textit{Finally}, Meta Pixel also supports advanced matching since 2016, where it collects certain information a user enters in web forms such as email, first name, last name, phone number, gender, birth date, city, state, zip, and country information \cite{Mutesix2016Advmatching, Meta2025Advmatching,senolLeakyFormsStudy2022}.

\vspace{0.05in} \noindent $\bullet$ \textbf{Tracking Restrictions:}
Meta Pixel has recently introduced tracking restrictions to limit the sharing of certain types of information with Meta. 
For example, Meta Pixel introduced \texttt{UnwantedData} configuration, which filters certain URL (or payload) parameters. 
The filtering is based on what Meta calls \texttt{sensitive} and/or \texttt{blacklisted} keys (Section \ref{sec:reverse-engineering-framework}).
In 2023, Meta introduced Core Setup that imposed more stricter restrictions on websites that belong to sensitive categories \cite{Hallen2023CoreSetup}. 
In Core Setup \cite{Meta2025CoreSetup}, Meta Pixel is supposed to not share custom parameters and the URL information is to be limited to the domain level.
While an advertiser can in theory enable Core Setup on its own, it is mostly turned on by Meta \cite{Reddit2024Coresetup, Hallen2023CoreSetup}.

% \vspace{-2mm}
\subsection{Meta Pixel Installation \& Configuration}
\label{sec:meta-pixel-configurations}

Figure~\ref{fig:loading-meta-pixel} shows how Meta Pixel is loaded on a website.
\noindent When an advertiser creates a pixel, Meta assigns it a unique identifier called Pixel ID.
Meta provides the base code that advertisers are recommended to include in the head section of their website.
When a user visits the advertiser's website, the base code executes and \textit{fbevents.js}~\cite{meta2017fbeventsfile}---common across all websites with Meta Pixel install- ed---is downloaded. 
The \textit{fbevents.js} script subsequently loads a configuration script that is specific to the Pixel ID. 
This configuration script reflects the information about various features and settings defined for the Pixel ID.
Once Meta Pixel loads, each event is collected via a HTTP GET/POST request at Meta’s \textit{https://www.facebook.com/ tr/?id=[PixelID]} server.

Our analysis shows that a portion of the configuration script is minified and obfuscated.
However, a portion towards the end of the configuration script, within the {\color{mygray}{fbq.\allowbreak registerPlugin(...)}} function, includes structured information related to adoption and configuration of various features. 
It contains three main sets of functions:\\

\noindent \textbf{(1) instance.optIn(<pixelID>, <configName>, <bool>)} opts in to the respective configuration for the Pixel ID as shown in the example below:
\begin{lstlisting}
instance.optIn("1234567891234567", "UnwantedData", true);
\end{lstlisting}
\label{listing:optin-unwanteddata}

\noindent \textbf{(2) config.set(<pixelID>, <configName>, <configJSON>)} allows setting the configuration at a finer granularity as dictated by the input JSON as shown in the example below:\\
    
\begin{lstlisting}
config.set("1234567891234567","unwantedData", {
     "blacklisted_keys": {
        "ViewContent": {
            "cd": ["em"], "url": ["lat", "lng"]
        },
    },
    "sensitive_keys": {
        "PageView": {
            "cd": ["d3857b12b4cea..."], // SHA-256 hash
        }
    }
});
\end{lstlisting}
\label{listing:config-unwanteddata}
% "url": []

\noindent \textbf{(3) fbq.set(<configurationName>, <pixelID>, <list>)} lists specific configurations on how a Pixel ID handles data collection and processing.
\begin{lstlisting}
fbq.set("estRules", "1234567891234567", [{
    "condition": { ...
        "conditions": [{ ...
            "value": "Submit my portfolio"
        }]
    },
    "derived_event_name": "SubmitApplication", ...
    "rule_id": "3690133590227007"
}]);
\end{lstlisting}
\label{listing:config-estrules}

% \vspace{-3mm}
\noindent This set of functions in the configuration script captures various features and their configurations. 
While additional configurations may be introduced over time, our analysis of the configuration scripts from 2017-2024 shows that the structure of these functions remains consistent.
Hence, we rely on the configuration script in developing our differential analysis based reverse-engineering framework in Section~\ref{sec:reverse-engineering-framework}.

% \vspace{-2.5mm}
\subsection{Related Work}

\textbf{Prevalence of tracking pixels.} 
The research community has extensively studied the prevalence of tracking pixels \cite{vekaria2025sok,alsaid2002detecting, martin2003hidden,Bekos2023HitchikerWebconf, Lerner2016InternetjonesUSENIX, Englehardt2016TrackingCCS, libert2015exposing, roesner2012detecting, Iqbal2021fingerprintingSP, vlajic2018online,vekaria2021differential}. 
Libert \cite{libert2015exposing} reported Google Analytics and Facebook Like button on 46\% and 21\% of the top-million websites, respectively. 
Englehardt and Narayanan \cite{Englehardt2016TrackingCCS} reported Google and Facebook's tracking pixels on 67\% and 24\% of top-1M websites, respectively.
Complementing this line of work, Lerner et al. \cite{Lerner2016InternetjonesUSENIX} conducted a longitudinal study of tracking pixels using data from the Internet Archive's Wayback Machine. 
The authors reported a steady increase in the prevalence, variety, and capabilities of tracking pixels between 1996 and 2016.
Ruohonen et al. \cite{ruohonen2018invisible} found 1x1 image pixels on 31\% of top-500 websites.
Fouad et al. \cite{fouad2018missedPETS} expanded the analysis beyond 1x1 image pixels and reported image pixels on 95\% of the crawled websites. 
The authors further classified image pixels based on their tracking behaviors.

\begin{table}[h!]
\centering
\small
\caption{Comparison of our study's scope and methodology against the most relevant prior works.}
\label{tab:related-work-comparison}
\renewcommand{\arraystretch}{1.15}
\begin{tabular}{l c c c c}
\toprule
\textbf{Analysis Aspect} 
& \makecell{\textbf{Our}\\\textbf{Work}} 
& \makecell{\textbf{Bekos et}\\\textbf{al. (2023)}} 
& \makecell{\textbf{Bekos et}\\\textbf{al. (2025)}} 
& \makecell{\textbf{Kieserman}\\\textbf{et al. (2025)}} \\
\midrule

\multicolumn{5}{l}{\textbf{Data Sources}} \\
Config. Script & \checkmark &  & \checkmark & \checkmark \\
Traffic/HTML &  & \checkmark &  & \checkmark \\

\addlinespace
\multicolumn{5}{l}{\textbf{Configurations}} \\
InferredEvents & \checkmark &  &  &  \\
Microdata & \checkmark &  &  &  \\
ESTRules & \checkmark &  &  &  \\
First-party Cookies & \checkmark &  &  &  \\
AAM & \checkmark &  & \checkmark & \checkmark \\
UnwantedData & \checkmark &  &  &  \\
CoreSetup & \checkmark &  &  &  \\

\addlinespace
\textbf{Longitudinal} & \checkmark &  &  &  \\

\bottomrule
\vspace{-8mm}
\end{tabular}
\end{table}

Recent studies \cite{Bekos2023HitchikerWebconf, amieur2024client,kieserman2025trackersnotequalPETS} have shifted focus towards analyzing specific tracking pixels in greater depth. 
The studies by Bekos et al. \cite{Bekos2023HitchikerWebconf, Bekos2025PIIxelCCS} (2023 and 2025) and Kieserman et al. \cite{kieserman2025trackersnotequalPETS} are most relevant to our work. 
Bekos et al. (2023) found that 23\% of top-10K websites use Meta Pixel and examined only events and first-party cookies using network traffic, without analyzing the full configuration script \cite{Bekos2023HitchikerWebconf}. 
Kieserman et al. later analyzed Meta Pixel's configuration scripts and network traffic across a sample of top-1M websites (28.2\% presence) and found that 62.3\% of these Pixels were configured to share information from form fields for Advanced Matching. 
Bekos et al. (2025) performed a similar analysis (18.7\% presence) and found Advanced Matching enabled on 65.45\% of these Pixels \cite{Bekos2025PIIxelCCS}.
Both of these studies focus exclusively on Advanced Matching as a mechanism for sharing PII from form fields.
In contrast, we study a \textit{broader} and \textit{previously unstudied} set of Meta Pixel configurations using a comprehensive reverse-engineering analysis of the configuration script. 
Our study includes the configurations analyzed in prior work, but also key ones that they do not examine, such as automatic events, the event setup tool, and tracking restriction mechanisms. 
We further extend prior work through a longitudinal analysis of all configurations via the Wayback Machine, enabling us to link regulatory and platform changes to tracking behavior on health websites. 
See Appendix Table \ref{tab:related-work-comparison} for a comparison of our study's scope and methodology with related work.

\noindent \textbf{Tracking pixels on health websites.}
The research community has also investigated the use of tracking pixels on potentially sensitive websites such as those related to health. 
Intuitively, tracking pixels on health websites pose different privacy risk as compared to those on a generic website. 
Health information is considered more sensitive \cite{ur2012smart,leon2013matters,melicher2016not} and observes greater protection under various laws and regulations (e.g., HIPAA \cite{Hipaa1996}). 
Zeng et al. \cite{zeng2025healthriskCHI} audited the outcome of such tracking, finding that every 100 health-related pages browsed resulted in 2.3 more health ads shown to a user's profile. 
Libert \cite{libert2015privacy} reported that 91\% of the 80,142 health-related web pages include one or more third-party, with 78\% for Google, 38\% for comScore, and 31\% for Facebook. 
McCoy et al. \cite{mccoy2020covidJAMA} found that 535 out of 538 COVID-19 related web pages included at least one third party.
Friedman et al. \cite{friedman2023hospitaltrackingHealthAffairs} analyzed 3,747 hospital websites and found that 98.5\% shared information with third parties, with 55.6\% and 98.5\% sharing information with Google and Meta, respectively.
Huo et al. \cite{huo2022all} reported that 67 and 7 out of the 459 patient portals include Google Analytics and Facebook Pixel, respectively. 
Markup's \textit{Pixel Hunt} project~\cite{markup2022pixelhunt} investigated the use of Meta Pixel across 100 hospital websites in the United States~\cite{markup2022medicalsensitive}. 
The academic and journalistic research into the use of tracking pixels on health websites has prompted regulatory action in the US. 
In 2022, the Federal Trade Commission (FTC)~\cite{ftc2025} issued enforcement actions against digital health platforms like \textit{GoodRx} and \textit{BetterHelp} for sharing sensitive health information due to their use of tracking technologies such as Google, Facebook, Snapchat, Criteo, and Pinterest \cite{boynton2023goodrx, khan2023BetterHelp}. 
In 2022, Department of Health and Human Services (HHS)~\cite{hhs2025ocr} released a bulletin~\cite{hhs2024HIPAAtracking} on the use of tracking technologies on health websites~\cite{alder2023warninghipaa}.
Sub-

\noindent sequently, in 2023, the FTC and HHS issued warning letters to 130 healthcare providers using tracking technologies on their websites~\cite{alder2023warninghipaa, FTC2023HeartTracking}.
Our longitudinal analysis sheds light on the impact of these enforcement actions on the adoption and configuration of tracking pixels on health websites.

%%

% \vspace{-3mm}
\section{PixelConfig}
\label{sec:reverse-engineering-framework}
\begin{figure}[t]
    \centering
    \begin{subfigure}[b]{1\linewidth}
        \centering
        \includegraphics[width=\linewidth]{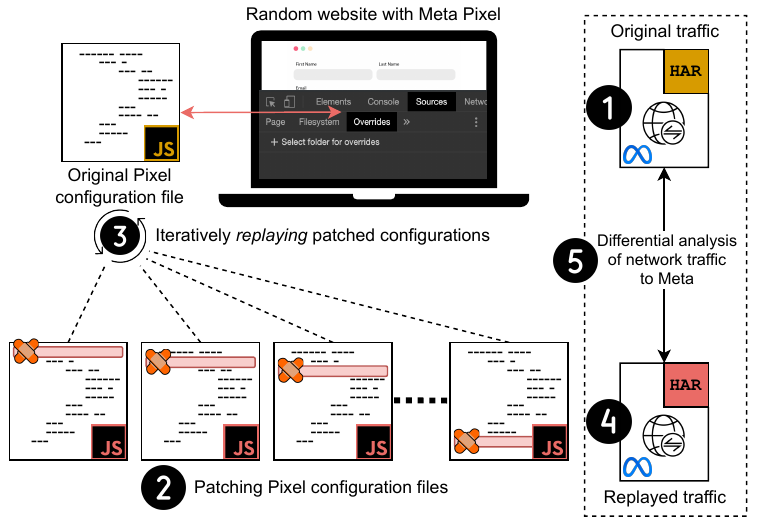}
        \vspace{-5mm}
        \caption{Replaying patched Pixel configuration scripts on a random website with Meta Pixel, followed by differential analysis of network traffic to Meta before and after replays.}
        \label{fig:reverse-enginerring-a}
        \vspace{1mm}
    \end{subfigure}
    \hfill
    \begin{subfigure}[b]{1\linewidth}
        \centering
        \includegraphics[width=\linewidth]{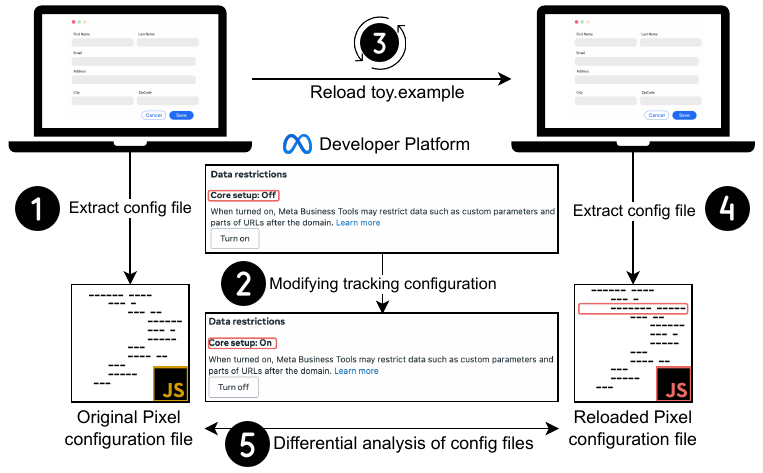}
        \vspace{-5mm}
        \caption{Comparing changes in client-side Pixel configuration scripts before and after changing feature settings on Meta Business Manager.} %  platform
        \label{fig:reverse-enginerring-b}
    \end{subfigure}
    \vspace{-6mm}
    \caption{\texttt{PixelConfig}: Our differential analysis based reverse-engineering framework comprising (a) \& (b)}
    % \caption{Our reverse-engineering framework comprising: (a) replaying of patched configuration scripts and (b) analyzing client-side configuration script changes resulting from changes in platform feature settings.}
    \label{fig:pixel-config}
    \vspace{-3mm}
\end{figure}

In this section we describe \texttt{PixelConfig}, a reverse-engineering framework to study the configurations of a Meta Pixel instance. 
\texttt{PixelConfig} involves a two-fold strategy as shown in Figure~\ref{fig:pixel-config}.
\textit{\textbf{First}}, we perform \textit{code patching} of different configurations defined in a Meta Pixel instance by iteratively commenting out or removing specific lines related to a given configuration. 
We replay the patched configuration script on client-side using the override functionality in Chrome DevTools~\cite{chrome-override}.
We inspect network traffic to Meta's servers (i.e., \textit{facebook.com}) to understand the effect of patched line in the configuration script.
\textbf{\textit{Second}}, we create a test website, sign-up for a Meta developer account, and iteratively create pixels to test the effect of different features and configuration settings available in Meta Business Manager.
We do this because changing various feature settings result in changes in the client-side configuration script of the Meta Pixel instance.
Thus, we perform differential analysis of network traffic to Meta's servers before and after replaying a patched configuration script as well as differential analysis of the configuration script before and after modifying various feature settings in Meta Business Manager.
Below we explain the how we use this differential analysis to ascertain the three categories of Meta Pixel features: activity tracking, identity tracking, and tracking restrictions.
Table~\ref{tab:config_mapping} summarizes the mapping.

\begin{table}[ht]
    \centering
    \caption{Meta Pixel features: Summary of Reverse-engineered configuration mapping, timeline, defaults.}
    \vspace{-2mm}
    \label{tab:config_mapping}
    \renewcommand{\arraystretch}{1.8} % increase row height for vertical centering
    % \small
    \fontsize{8.5}{11}\selectfont
    \begin{tabular}{@{}c c c c@{}}
    % \begin{tabular}{@{}m{1cm} m{2.5cm} m{2.5cm} m{1.2cm}@{}}
        \toprule                            
        \textbf{\makecell{Tracking\\Category}} & \textbf{\makecell{Tracking\\Feature\\(Start Year)}} & 
        \textbf{\makecell{Pixel\\Configuration\\(Since Year)}} &   \textbf{\makecell{Default\\Status}} \\
        \midrule
    
        \multirow{2}{*}{\rotatebox[origin=c]{90}{\parbox[c]{5em}{\centering \textbf{Activity}}}}
        & \makecell{Automatic Events\\(2017)}  & \makecell{AutomaticSetup \\ InferredEvents\\(2017)} & \textcolor{customgreen}{\textbf{Enabled}} \\
        \cmidrule{2-4}
        & \makecell{Event Setup Tool\\(2019)} & \makecell{ESTRuleEngine\\(2023)} & \textcolor{customred}{\textbf{Disabled}}\ \\

        \midrule

        \multirow{2}{*}{\rotatebox[origin=c]{90}{\parbox[c]{5em}{\centering \textbf{Identity}}}}
        & \makecell{First-Party Cookies\\(2018)} & \makecell{FirstPartyCookies\\(2018)}  & \textcolor{customgreen}{\textbf{Enabled}} \\
        \cmidrule{2-4}
        & \makecell{Auto. Adv. Matching\\(2018)} & \makecell{AutomaticMatching\\(2018)} & \textcolor{customred}{\textbf{Disabled}}\\

        \midrule

        \multirow{2}{*}{\rotatebox[origin=c]{90}{\parbox[c]{6em}{\centering \hspace{1mm}\textbf{ Restrictions}}}}
        & \makecell{Prohibited Data Filter\\(Unknown)} & \makecell{UnwantedData\\(2020)} & \textcolor{customred}{\textbf{Disabled}} \\
        \cmidrule{2-4}
        & \makecell{Core Setup\\(2024)} & \makecell{ProtectedDataMode\\(2023)} \ & \textcolor{customred}{\textbf{Disabled}} \\

        \bottomrule
    \end{tabular}

    \vspace{1ex}
    \raggedright
    % \footnotesize
    % \textsuperscript{+}
    % \textsuperscript{+}\, While we find \texttt{InferredEvents} to be configured by default, our analysis suggests \texttt{AutomaticSetup} has been deprecated across websites (Section \ref{sec:archaeology-activty}). A developer can disable Automatic Events by including {\color{mygray}\texttt{fbq(`set',`autoConfig','false','PixelID)}}\cite{meta2025Pixeladvanceddoc} as part of the base code.\\
   
    \vspace{-3mm}
\end{table}

% \vspace{-2mm}
%%
\subsection{Activity Tracking}
\label{sec:reverse-eng-activity}

\noindent {\textbf{Automatic Events.}}
Meta Pixel implements the default \texttt{PageView} event along with the automatic \texttt{Microdata} and \texttt{SubscribedButton- Click} events. 
While the \texttt{PageView} event is defined within the base Pixel code, the automatic \texttt{Microdata} and \texttt{SubscribedButtonClick} events are not. 
%
% To examine their configuration, 
We patch the \texttt{AutomaticSetup} configuration by removing the call to {\color{mygray}\texttt{instance.optIn(<PixelID>, "AutomaticSet- up",true)}}.
Upon replaying the patched pixel, the \texttt{Microdata} event does not fire when the page is loaded.
Similarly, when we patch both the \texttt{AutomaticSetup} and \texttt{InferredEvents} configurations from the Pixel configuration script by removing their respective {\color{mygray}\texttt{instance.optIn}} calls and replay the patched script, the \texttt{Subscrib- edButtonClick} event no longer fires when a button is clicked. 
Both \texttt{Microdata} and \texttt{SubscribedButtonClick} events are enabled by default due to the presence of the \texttt{AutomaticSetup} and \texttt{InferredEve- nts} configurations in the Pixel configuration script. 
However, the \texttt{AutomaticSetup} configuration no longer appears in the script for newly created pixels, suggesting either its deprecation or consolidation into the \texttt{InferredEvents} configuration.
%
% It appears to have been deprecated or consolidated into the \texttt{InferredEvents} configuration. 

%%
\noindent {\textbf{Event Setup Tool.}}
In addition to automatic events, advertisers can configure standard and custom events that can be triggered by specific user actions such as button clicks.
To investigate how such events are set up, 
we use the Event Setup Tool in Meta's Business Manager to configure a new event triggered by a button click. % platform
Comparing the configuration script before and after configuring the event, we observe that only the \texttt{estRules} object is updated to include a new object representing the newly configured event. 
Next, by patching and replaying the script after removing the {\color{mygray}{fbq.set("estRules", ...)}} call from Listing 3 (\Cref{sec:meta-pixel-configurations}) on our test website, we find that the \texttt{Submit Application} event no longer fires, proving that the \texttt{derived\_event\_name} key within this object defines the triggered event type.
We also observe that the button's inner HTML matched the `value' field within the \texttt{condition} object of the disabled \texttt{estRules}, which, in this case, was ``Submit my portfolio.''

%%
% \vspace{-2mm}
\subsection{Identity Tracking}
\label{sec:reverse-eng-identity}

\noindent {\textbf{First Party Cookies.}}
First-party cookies are used to track a user across different sessions and link interactions back to a specific device or session, especially when third-party cookies are blocked. 
By default, first-party cookies are enabled when a Meta Pixel is installed---transmitting \texttt{\_fbp} and \texttt{\_fbc} cookies as parameters with each event. 
To examine how the inclusion of first-party cookies is controlled, we configure a Meta Pixel instance on a test website with first-party cookies initially disabled.
Upon enabling first-party cookies through the Events Manager, we observe an addition of the {\color{mygray}{instance.optIn(<Pixel ID>, "FirstPartyCookies", true)}} call in the configuration script.
Replaying the pixel with this configuration results in the inclusion of both \texttt{\_fbp} and \texttt{\_fbc} values in the event payload. 
Conversely, patching and replaying the Pixel by removing the {\color{mygray}{instance.optIn}} call prevents the transmission of the two values.

\noindent{\textbf{Automatic Advanced Matching (AAM).}}
AAM enables advertisers to share hashed versions of user information (Section \ref{sec:how-pixel-tracks-users}). 
When AAM is enabled (it is not enabled by default), Meta Pixel automatically detects different input types when a user submits a form and shares the hashed versions of these identifiers as \texttt{udff[<parameter- Key>]} parameters in the payload of the \texttt{SubscribedButtonClick} event.
For instance, the hashed email address is transmitted as \texttt{udff[<em>]}, the hashed phone number as \texttt{udff[<ph>]}, and so on.
Enabling AAM in the Meta Events Manager for our test website results in the inclusion of both {\color{mygray}{instance.optIn(<Pixel ID>, "AutomaticMatching", true)}} and {\color{mygray}{config.set (<PixelID>, "automaticMatching", \{...\})}} calls in the Pixel configuration script.
The {\color{mygray}{config.set}} call also includes a \texttt{selectedMatchKeys} array that specifies which user information fields to track:
\begin{lstlisting}
config.set("1286678629287552", "automaticMatching", { 
    "selectedMatchKeys": [
        "em", "ph", "fn", "ln", "ge", "db", 
        "ct", "st", "zp", "country", "external_id"
    ]
});
\end{lstlisting}
This array contains keys (e.g., \texttt{em} for email) corresponding to the user data fields that the advertiser chooses to track. 
Removing a key from the \texttt{selectedMatchKeys} array prevents transmission of the corresponding hashed user information.
Moreover, patching and replaying the Pixel after removing {\color{mygray}{instance.optIn}} call prevented the hashed user data from being sent, confirming that AAM is disabled.

\vspace{-2mm}
\subsection{Tracking Restrictions}
\label{sec:reverse-eng-controls}

\noindent {\textbf{Unwanted Data.}}
The \texttt{UnwantedData} configuration defines rules that limit the sharing of specific parameters with Meta. 
These rules are categorized into \texttt{blacklisted\_keys} and \texttt{sensitive\_keys}, and are applied at the event level (e.g., \texttt{PageView}, \texttt{CompleteRegistrat- ion}, etc.). 
Each event type can have rules for both custom data parameters (\texttt{cd}) and URL query parameters (\texttt{url}). 
The technical distinction between \texttt{blacklisted\_keys} and \texttt{sensitive\_keys} lies in how parameters are specified in the \texttt{UnwantedData} configuration.
\texttt{blacklisted\_keys} lists parameter names in plain text (e.g., ["\texttt{lat}", "\texttt{lng}"]).
\texttt{sensitive\_keys} consists of SHA-256-hashed parameter names (e.g., \texttt{d3857b12b4cea...} for a hashed parameter name as shown in Section \ref{sec:meta-pixel-configurations}). 
For instance, in the payload of a \texttt{ViewContent} event triggered on the URL \textit{\url{https://www.example.com?lat=40.00\&lng=35.00}}, the \texttt{dl} parameter is sanitized to \textit{\url{https://www.example.com?lat=\_removed\_\&lng=\_removed\_}} if \texttt{lat} and \texttt{lng} are either listed as \texttt{blacklisted\_ keys} or their SHA56 hashes are included in \texttt{sensitive\_keys}. 
We confirm this by removing a parameter from the \texttt{UnwantedData} configuration for a specific event type (e.g., excluding \texttt{lat} from \texttt{ViewContent}'s \texttt{blacklisted\_keys}), which was responsible for transmission of the parameter. 
Conversely, adding a parameter's SHA256 hash to \texttt{sensitive\_keys} triggered its sanitization. 
For custom data parameters, including a parameter name in the
\texttt{cd} array under \texttt{blacklisted\_keys} or its hash under \texttt{sensitive\_keys} suppresses its transmission entirely. 
Next, we configured an event on a test website to share \texttt{dob} as a custom data parameter to Meta.
Upon triggering the event, a notification in Meta Events Manager stated that the \texttt{dob} parameter was ``Blocked by Meta'' (Figure \ref{fig:arch-dob-blocked}). 
Subsequent inspection of the downloaded Pixel configuration script revealed that the \texttt{cd} array in the \texttt{blacklisted\_keys} had been auto-updated to include \texttt{dob}.

\noindent {\textbf{Core Setup.}}
Core Setup is a configuration that enforces strict restrictions on the sharing of custom parameters and URL query parameters.
When a website is placed under Core Setup, indicated by the {\color{mygray}{instance.optIn(<Pixel ID>, "ProtectedDataMode," true)}} call, the \texttt{cd} parameters are omitted, and \texttt{dl} and \texttt{rl} parameters which typically include full URLs and referrer URLs, are truncated to only the domain.
To validate this behavior, we configure a Meta Pixel on a test website and enable Core Setup via the developer platform.
Upon enabling Core Setup, we observe an addition of the {\color{mygray}{instance.optIn(<Pixel ID>, "ProtectedDataMode," true)}} call in the configuration script.
Patching and replaying the Pixel after removing the {\color{mygray}{instance.optIn}} call reverted the behavior, confirming that \texttt{ProtectedDataMode} reflects the Core Setup configuration. 
Our controlled testing on the test site aligns with our observations on other sites (e.g. \textit{wexnermedical.osu.edu}), where toggling Core Setup (via {\color{mygray}{instance.optIn(<Pixel ID>, "ProtectedDataMode," true)}} call), consistently modifies the payload in the specified manner.

%%

%%
% \vspace{-2mm}
\section{Crawling Methodology}
\label{sec:methodology}
Here, we explain the selection of websites (\Cref{sec:website-selection}), collection of their longitudinal snapshots (\Cref{sec:snapshot-crawling}), and crawling of Meta Pixel configuration scripts (\Cref{sec:extracting-pixelids}).

%%
% \vspace{-2mm}
\subsection{Website Curation} 
\label{sec:website-selection}
We curate US-focused health websites from two sources -- American Hospital Association (AHA)~\cite{aha-website} and Centers for Medicare and Medicaid Services (CMS)~\cite{cms}. 
We obtained 5,685 AHA member hospital websites from AHA DataQuery \cite{aha2024dataquery}.
CMS provides a public dataset of 115,646 providers in the USA~\cite{CMS2025data}. 
We identify 53,432 unique active providers. 
Unlike AHA data, CMS dataset does not contain website information so we rely on Google Search (``[$name$] [$city$] [$state$] official website'') to identify the providers' websites. 
Overall, we identify 18,327 unique US-focused health websites -- 3,272 websites from AHA and 15,055 websites from CMS.
To compare health websites against a baseline, we use top-10K websites (Tranco ID: G672K)~\cite{Pochat2019TrancoNDSS} as a control group. 
We limited the analysis to top-10K websites due to the crawling limitations of the Wayback Machine as explained below. A total of 42 websites were common in both the top-10K and the health websites. 
Our analysis of the website categories within the top-10K dataset reveals that a substantial proportion of sites belong to Business, Clothing, and Technology-related industries (see Appendix~\ref{sec:website-categorization}).

\vspace{-1mm}
\subsection{Crawling Website Snapshots}
\label{sec:snapshot-crawling}
We rely on Internet Archive's Wayback Machine~\cite{archive2025Waybackmachine} for longitudinal analysis of tracking pixels on our set of health and control websites.
Wayback archives websites and their resources (e.g., scripts, images). 
It has already archived more than 900 billion web pages since 1996.

We begin by crawling a website's snapshot on the Wayback Machine to detect and extract installed Meta Pixel IDs.
We use the CDX Server API~\cite{archive2025CDXapi} to collect historical website snapshots available on the Wayback Machine. 
The CDX records provide metadata about available snapshots, including timestamps, which allow us to construct appropriate Wayback Machine URLs for each snapshot.
The CDX Server API was queried in batches of up to 100,000 records per request for each website, retrieving website snapshot records from 2017 onwards. 
For the majority of websites, a single request was sufficient to retrieve all available records. 
By limiting the number of CDX API retries to five, snapshot timestamps obtained from the records were used to generate corresponding \url{archive.org} URLs of website snapshots.
For example, URL for the Wayback’s snapshot of \textit{\url{https://www.facebook.com/}} dated Sept 1, 2024, is {\textit{\url{https://web.archive.org/web/20240901000408/https://www.facebook.com/}}}.

Due to Wayback Machine’s rate limits and inconsistent archiving of internal pages, we limited snapshot retrieval to twice per year (January 1 and July 1) for each website and focused only on landing pages.
In absence of a website snapshot on January 1 or July 1, we search for the closest archived snapshot.
Each archived website snapshot was crawled using a Selenium \cite{selenium2025} driven Chrome browser. 
%
% When using 
Selenium's default page load strategy, ensures that each webpage is fully loaded during the crawl, allowing dynamic elements to be rendered in the HTML snapshot~\cite{selenium2025options}. 
Moreover, given that website crawling on the Wayback Machine can face issues, we retry each archived website snapshot ten times.

%%
% \vspace{-1mm}
\subsection{Extracting Pixel ID and Configuration}
\label{sec:extracting-pixelids}
To identify Meta's unique Pixel IDs for the current website, we parse its HTML snapshots to locate the script tag that loads the configuration script and the initialization of the {\color{mygray}{fbq()}} function.
Once extracted, these Pixel IDs are used to identify and crawl the corresponding configuration scripts.
Thus, similar to how website snapshots are crawled, the CDX Server API is used to identify snapshots of Pixel configuration scripts by fetching the URL \textit{\url{https://connect.facebook.net/signals/config/<PixelID>}} as a prefix. For instance, a valid snapshot of the configuration script for \textit{example.com} with Pixel ID 1234 as of August 1, 2024, corresponds to the URL \textit{https://web.archive.org/web/20240801000502/https://connect.facebook .net/signals/config/1234...}
Unlike the website snapshots (crawled using Selenium), the identified configuration script snapshots are directly fetched using the Python's requests library to minimize overhead. 
To ensure accuracy, each configuration script is assigned to a website only for the years in which that Pixel ID was observed in the website snapshot. For instance, a configuration script for Pixel ID `A' archived in 2019 is only associated with a website for 2019 if that website's 2019 snapshot contained Pixel ID `A'.

\begin{figure}[t] 
    \centering
    \includegraphics[width=\linewidth]{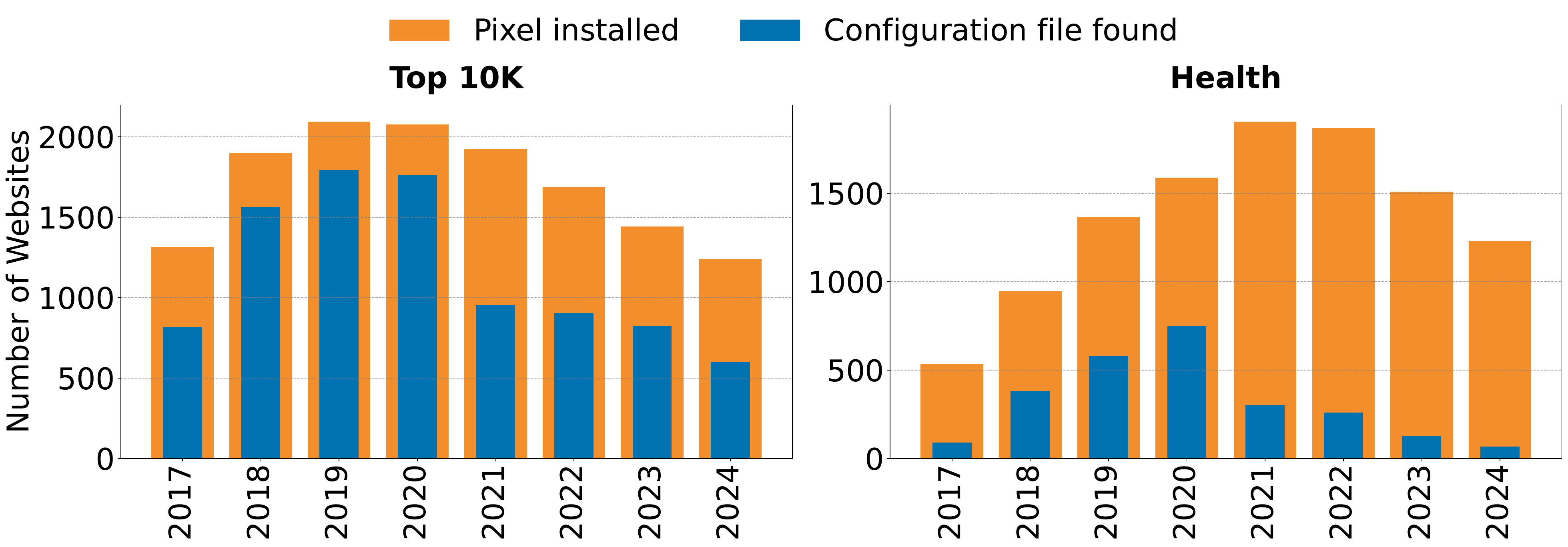}
    \vspace{-4mm}
    \caption{Number of control and health websites with Meta Pixel (\textcolor{orange}{orange}) and number of websites with configuration script in that year (\textcolor{myblue}{blue}) from 2017-2024.}
    % \caption{Number of Top 10K (left) and Health (right) websites with Meta Pixel installed (\textcolor{orange}{orange}) and corresponding number of websites where configuration script was also found that year (\textcolor{myblue}{blue}) from 2017-2024.}
    \label{fig:pixel-presence}
    \vspace{-4mm}
\end{figure}

Figure ~\ref{fig:pixel-presence} illustrates the temporal distribution of websites with Meta Pixel installations and those for which configuration scripts were archived. 
Out of all the control and health websites, we find 3,263 and 2,917 websites, respectively, with at least one snapshot containing Meta Pixel installation across the years. 
The other websites in these groups showed no evidence of Pixel installation based on their HTML.
However, not all of these Pixel installations had corresponding configuration scripts archived by the Wayback Machine.
Specifically, we find configuration scripts for 2,656 control and 1,159 health websites, indicating that while a Pixel was present, the associated configuration script was not always archived.
This inconsistency in the Wayback Machine's archival coverage means that our longitudinal comparison across years may not comprise the same set of websites, as some configuration scripts are not consistently archived once every year.
Nevertheless, the number of configuration scripts per year remain sufficient for trend analysis \cite{Lerner2016InternetjonesUSENIX}. 
Moreover, the trend of declining Pixel adoption across top-10K websites in recent years is consistent with independent external measurements \cite{Builtwith2025Metapixel}. 
Although the absolute numbers may represent a conservative estimate due to archival limitations, we eliminate any potential bias stemming from these limitations by performing a series of validation experiments (details in Section~\Cref{subsec:validation}).

% \vspace{-4mm}
\subsection{Archival Validation}
\label{subsec:validation}
To assess the completeness of the Wayback Machine's archival process for Pixel installation and their configurations, we conducted a validation study by crawling the live version and the most recent Wayback Machine snapshot for randomly sampled 1,000 top-10K and 1,000 health websites. 
We restrict the sample to 1,000 websites per category to allow for comprehensive manual validation.

\noindent {\textbf{Pixel Archival Completeness.}} 
On live websites, we identified the Meta Pixel on 101 Tranco and 72 health domains. 
In contrast, the corresponding Wayback Machine snapshots contained the Pixel on 50 and 43 of these sites, indicating that approximately 51\% and 40\% of the live Pixel installations did not appear in the archived snapshots.
Manual analysis revealed several potential causes.
8 of the 80 missed installations corresponded to pages that loaded with error messages (e.g., Access Denied, 504 Gateway Timeout). 
In four additional cases, only the Wayback Machine’s wrapper page (including the toolbar) was archived, suggesting that the underlying webpage failed to load during capture. 
However, note that our sampling frequency of twice per year mitigates missed installations due to such archival limitations.
Finally, since none of the analyzed websites’ \texttt{robots.txt} files explicitly disallowed common Wayback user agents (e.g., \texttt{ia\_archiver}), these discrepancies likely stem from limitations in the Wayback Machine’s archiving process rather than website crawling restrictions.

\noindent {\textbf{Configuration File Availability.}} 
We also validated the availability of the corresponding configuration scripts. 
Using the Pixel IDs extracted from live websites, we queried the Wayback Machine for archived configurations. 
We found that 68.3\% of Pixels on live top-10k websites to have a configuration file archived within one year of the live crawl, compared to only 18.1\% for health websites. 
The disparity may be attributed to Wayback Machine's bias towards more frequent crawling and more complete archival of popular sites~\cite{tranos2019IntrnetAvaialabilitycompenvurbsys}, as their crawlers follow a snowball-like process that relies on high backlink volumes to discover and capture content.

\noindent {\textbf{Robustness of Findings.}} 
To ensure that these archival limitations do not undermine our longitudinal findings, we performed two robustness checks. 
First, to assess whether missing configurations skew our findings, we conducted a comparative analysis of Tranco and health websites sampled earlier (i.e., 1,000 each) for all configurations using their live configuration scripts, all of which were readily available. 
Second, to assess whether having a different set of ``websites with configurations'' across the years impacts comparability, we conducted the longitudinal analysis for all configurations on a stable set of top-10K and health websites for which data was consistently available for at least 4 out of the 8 years we study. 
In both cases, the observed trends (detailed in Appendix~\ref{sec:robustness-check}) were consistent with our primary results, ensuring robustness of our conclusions.

\vspace{-1mm}
% \section{Analysis of Meta Pixel Configurations}
\section{Analysis of Meta Pixel Configs}
\label{sec:archaeology}
Using the \texttt{PixelConfig} framework described in Section~\ref{sec:reverse-engineering-framework}, we map different Meta Pixel configurations to their corresponding tracking behaviors to assess their capabilities. 
Leveraging this framework, we conduct an archaeological analysis of Meta Pixel configurations on health and control websites from 2017 through 2024.
In this section, we investigate differences in configuration patterns, privacy implications, and ownership responsibilities across activity tracking, identity tracking, and tracking restrictions.
The following analysis of Meta Pixel configurations is performed on the subset of websites for which we successfully retrieved a configuration script in a given year (Figure \ref{fig:pixel-presence}). The exact website counts for both detected pixels and retrieved configurations are provided in Appendix (Table~\ref{tab:appendix-website-counts-basic}). A configuration is considered active in a given year if at least one of the site’s pixels exhibit that configuration at any point during that year.

\begin{figure}[t] % htbp 
    \centering
    \includegraphics[width=0.96\linewidth]{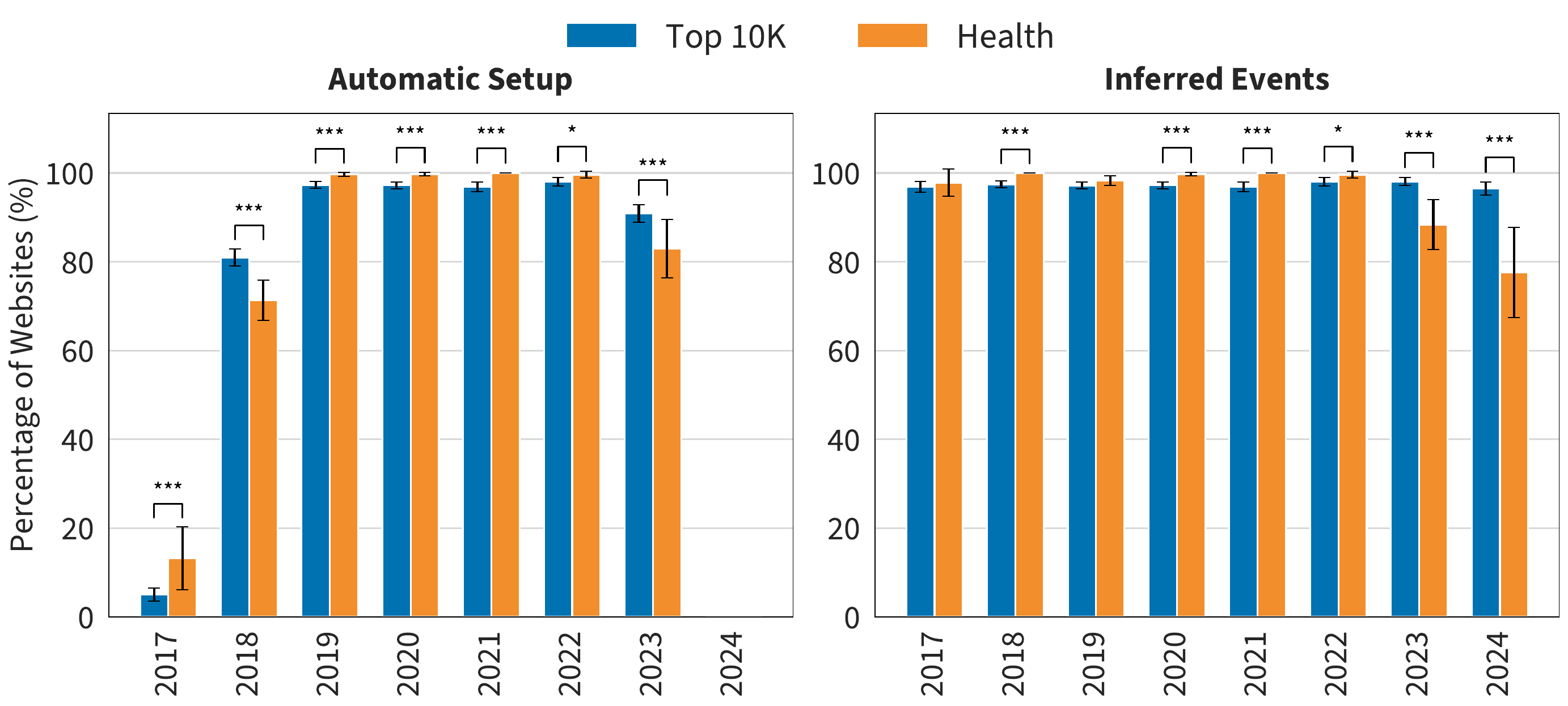} 
    \vspace{-4mm}
    \caption{Comparing percentage of \textcolor{myblue}{control} and \textcolor{orange}{health} websites (websites with configurations) using \texttt{AutomaticSetup} and \texttt{InferredEvents} by year from 2017 to 2024. Error bars represent 95\% confidence intervals. Asterisks indicate statistical significance determined by a two-proportion z-test (*$p$ < 0.1, ***$p$ < 0.01). Effect sizes for significant results, measured using Cohen’s $h$, ranged from $h_{0.21}$ to $h_{0.36}$ for \texttt{AutomaticSetup} and from $h_{0.23}$ to $h_{0.61}$ for \texttt{InferredEvents}.}
    \label{fig:automatic_setup}
    \vspace{-6mm}
\end{figure}

\subsection{Activity Tracking}
\label{sec:archaeology-activty}

% \yash{@Abdullah, please report all percentages upto one decimal point everywhere in Section 5.1, everywhere else in the paper, they seem to be upto 1 decimal. Also, ensure, they match the numbers reported in abstract, intro, and conclusion.}

\noindent {\textbf{Automatic Events.}} 
We saw that the \texttt{AutomaticSetup} configuration enables \texttt{Microdata} event, while the presence of either the \texttt{AutomaticSetup} or \texttt{InferredEvents} configurations enable the \texttt{SubscribedButtonClick}.
Figure \ref{fig:automatic_setup} (left) shows that the adoption of the \texttt{AutomaticSetup} configuration increased rapidly after its introduction in April 2017, becoming nearly ubiquitous by 2019. 
From 2019 to 2022, the configuration was present on over 97.2\% of control websites and over 99.6\% of health websites.
Across both categories combined, the adoption peaked at 98.4\% in 2022.
However, in 2023, the adoption dropped to 90.8\% for control and 83.0\% for health websites. 
Thus, we observe a decline in the use of the \texttt{AutomaticSetup} configuration in 2023 and do not observe it in any 2024 snapshots. 
While alternative explanations, such as archival bias, cannot be fully ruled out, a likely explanation is a platform-wide deprecation of \texttt{AutomaticSetup} by Meta.
This could be because \texttt{InferredEvents} is sufficient to trigger the \texttt{SubscribedButtonClick} event  and Meta could instead be gathering \texttt{Microdata} directly by crawling websites or through other mechanisms \cite{meta2025microdatadoc, meta2025WebCrawlers}.

Figure~\ref{fig:automatic_setup} (right) shows the adoption trends for the \texttt{InferredEvent} configuration. 
%
% Similar to \texttt{AutomaticSetup}, 
Its prevalence remained consistently high through 2017-2022, ranging from 97.1\% to 97.0\% for control websites and 97.8\% to 100\% for health websites.
In 2023, the adoption of \texttt{Inferred Events} declined among health websites (88.4\%), and further decreased in 2024 (77.6\%).
Our analysis reveals that 95\% of health websites and 85\% of control websites that no longer use the \texttt{InferredEv- ents} configuration were placed in Core Setup starting in July 2023.
Based on this observation, we speculate that Meta may have begun disabling the \texttt{SubscribedButtonClick} event under Core Setup automatically to mitigate the collection of potentially sensitive information, particularly for health websites.
% This suggests that Meta may have begun automatically disabling the \texttt{SubscribedButtonClick} event under Core Setup to mitigate the collection of potentially sensitive information, particularly for health websites.
%
% Unlike standard events discussed earlier, Meta  transmits button text as part of the \texttt{SubscribedButtonClick} event to help it understand the context surrounding the event. 
%
% Due to this, any user click could reveal the user's intent to Meta. For instance, a user navigating for a specific treatment by clicking on sections of website named as ``'colon cancer screening care'' or ``fertility and conception'' (as observed in \textit{healthgrades.com}) could disclose a user's PHI without a developer explicitly needing to set up any event.

\begin{figure*}[ht]
    \centering
    \begin{minipage}{\linewidth}
        \centering
        \includegraphics[width=\linewidth]{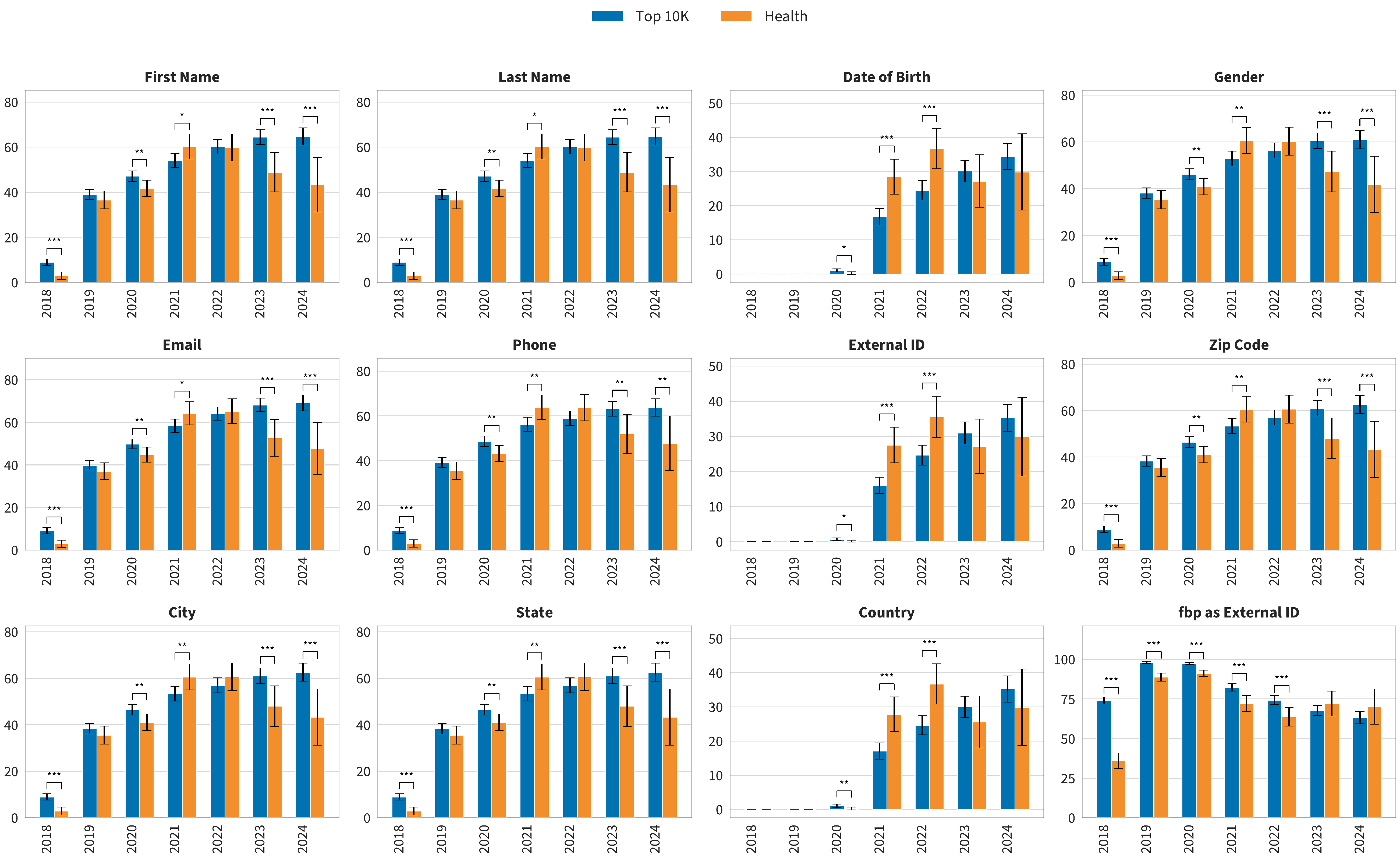}
        \vspace{-7mm}
        \caption{Percentage of \textcolor{myblue}{control} and \textcolor{orange}{health} websites with specific match keys configured for Automatic Advanced Matching (AAM) from 2018 to 2024. The plots show longitudinal trends for First Name, Last Name, Date of Birth, Gender, Email, Phone, External ID, Zip Code, City, State, Country, and instances where the fbp cookie is used as an external\_id (cases where first-party cookies are enabled but external\_id through AAM is not). Error bars represent 95\% confidence intervals. Asterisks indicate statistical significance determined by a two-proportion z-test (*$p$ < 0.05, **$p$ < 0.01, ***$p$ < 0.001).}
        \label{fig:arch-combined-aam}
        \vspace{-2mm}
    \end{minipage}
\end{figure*}

\noindent {\textbf{Event Setup Tool.}}
We detected that the \texttt{estRules} object captures standard and custom event configurations set up through Meta's Event Setup Tool.
Meta introduced its Event Setup Tool in April 2019~\cite{Fraser2019EST_2019}, but the \texttt{estRules} object does not appear in configuration scripts until January 2023.
Since its introduction, we observe that 45.3\% of health websites and 45.1\% of control websites have configured events using the Event Setup Tool.

\begin{figure}[t] % htbp
    \centering
    \includegraphics[width=0.96\linewidth]{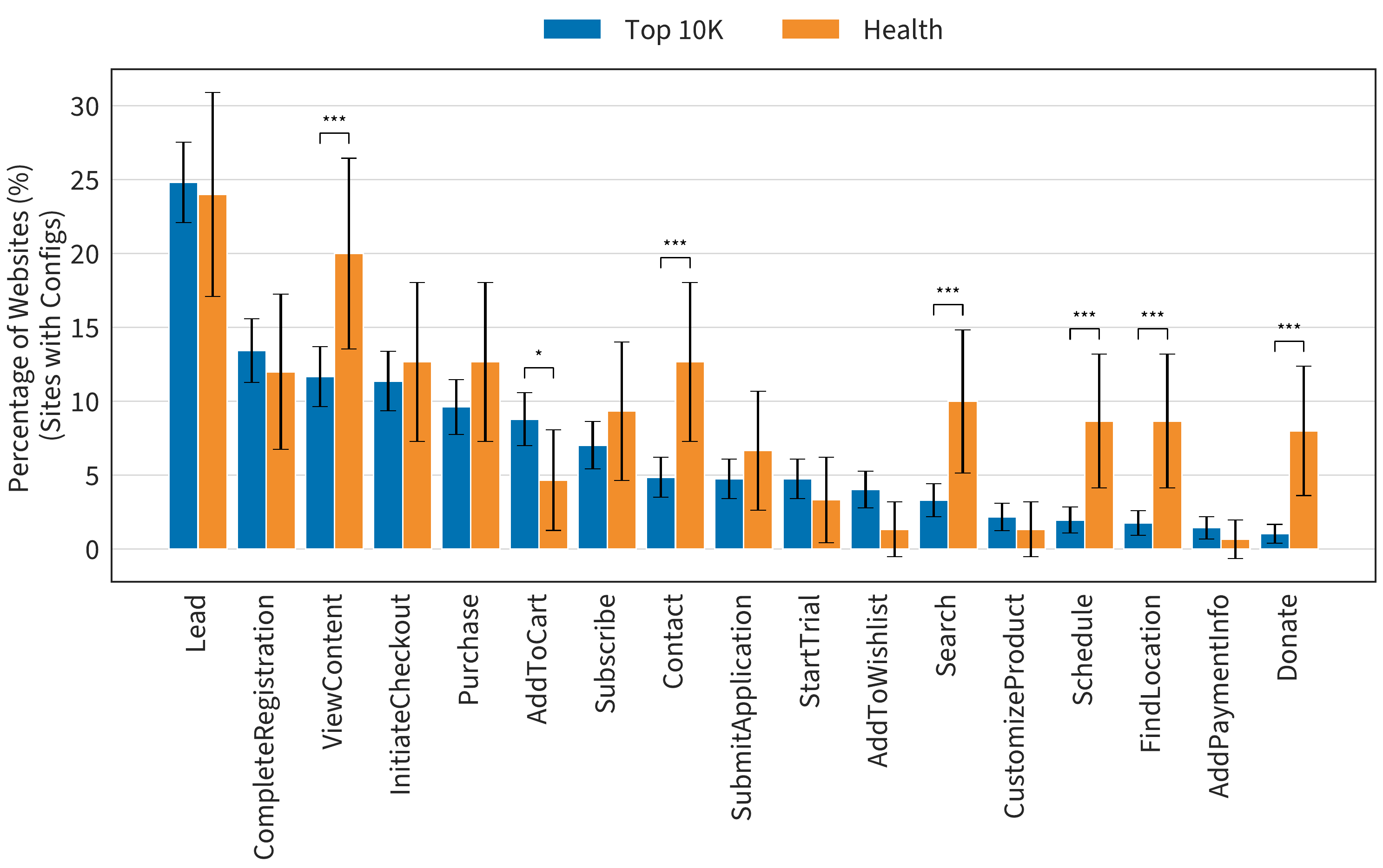} % Adjust width as needed (e.g., 0.9\linewidth)
    \vspace{-3mm}
    \caption{Percentage of websites configuring 17 standard events in 2023-2024 across \textcolor{myblue}{control} (N=967) and \textcolor{orange}{health} (N=150). These events are configured using Meta's Event Setup Tool and reflected in the \texttt{estRules} object. Asterisks indicate statistical significance determined by a two-proportion z-test (*$p$ < 0.1, ***$p$ < 0.01). Effect sizes for significant results, measured using Cohen’s $h$, ranged from $h_{0.17}$ to $h_{0.37}$.}
    \label{fig:standard-events}
    \vspace{-4mm}
\end{figure}

Figure~\ref{fig:standard-events} illustrates the distribution of standard events configured through \texttt{estRules} across control and health websites in 2023-2024.
The most commonly tracked event across both website groups is the \texttt{Lead} event, configured by approximately 25\% of websites.
This suggests that generating new leads is a primary focus for websites using the Event Setup tool.
However, certain events are configured disproportionately on health websites, aligning with user interactions typical of healthcare platforms.
For instance, health websites track the \texttt{Schedule} event to monitor patients booking appointments, the \texttt{Search} event to identify users seeking health-related information, and the \texttt{FindLocation} event to track searches for medical providers or facilities.
Additionally, \texttt{Donate} events are configured on health websites to track interactions with fundraising initiatives, such as donations to medical centers or research institutions.
Events such as \texttt{ViewContent} and \texttt{Contact} are also more prevalent on health websites.
Such events on health sites can result in collection of potentially sensitive health information by Meta.

Events configured through the Event Setup Tool are assigned a unique \texttt{rule\_id}, which Meta can use to link specific user interactions to button text or values defined in the estRules configuration (see Listing 5 in Section \ref{sec:meta-pixel-configurations}).
Since \texttt{rule\_id} is collected for each event, Meta can map user interactions with specific buttons, revealing contextually sensitive content on the webpage captured via the button text.
%
% For instance, we observe \textit{lakeareamc.org} (a healthcare provider) tracking a \texttt{Schedule} event when a user interacts with a button labeled ``request lung nodule screening appointment.''
%
For instance, we observe \textit{healthgrades.com} (a healthcare provider directory) tracks users clicking on buttons labeled as ``hiv'', ``hpv and genital warts'', ``sexual health'', ``schizophrenia'', and ``autism'' with a \texttt{ViewContent} event, and \textit{docasap.com} (a healthcare appointment scheduling service provider) tracks \texttt{Lead} events when their users navigate to certain sections on the website using button clicks such as ``birth control pills'', ``menstrual care products'', and ``erectile dysfunction.''
Another concerning example involves \textit{equitashealth.com}, an LGBTQ+ focused healthcare provider, which tracks HIV appointment registrations using the \texttt{CompleteRegistration} event when a user clicks a button labeled ``schedule your hiv sti testing appointment now''. 
These examples illustrate that Meta collects potentially sensitive health information using events configured using the Event Setup tool.

%%
% \vspace{-2mm}
\subsection{Identity Tracking}
\label{sec:archaeology-identity}

\noindent {\textbf{First-Party Cookies.}} 
First-party cookies can be used for both same-site tracking via the \texttt{\_fbp} cookie and cross-site tracking via the \texttt{\_fbc} cookie, which stores the \texttt{fbclid} click identifier.
On health websites, if a user performs actions that trigger event requests containing hashed identifiers (e.g., email, phone number) alongside the \texttt{\_fbp} cookie, Meta can link previous user activities to their Facebook identity.
In contrast, the \texttt{\_fbc} cookie captures the \texttt{fbclid} click identifier when a user clicks on an external link on Facebook, also enabling Meta to link a user's activities on an external website to their Facebook identity.

%%
% \noindent Meta leverages dark patterns to encourage advertisers to enable first-party cookies in Meta Pixel.
\noindent Meta leverages design choices that align well with established ``dark pattern'' taxonomies \cite{gray2024darkpatternsCHI}, encouraging advertisers to enable first-party cookies in Meta Pixel.
By default, first-party cookies are enabled when an advertiser creates a Meta Pixel, a practice that aligns with the \textit{Bad Defaults} dark pattern, in which the less privacy-protective option is set as the default. 
In contrast, disabling first-party cookies is notably more complex than toggling other features (such as Automatic Advanced Matching), requiring multiple steps that align with the \textit{Privacy Maze} dark pattern.
Advertisers must first click `Edit', then toggle off first-party cookies, and finally confirm the changes by saving (Figure \ref{fig:arch-firstpartycookies-toggle}).
Besides, Meta nudges advertisers in the Events Manager to keep first-party cookies enabled by promoting them as a mechanism to ``help deliver relevant ads to people who may be interested in your products'' (Figure \ref{fig:arch-firstpartycookies-toggle}).

\begin{figure}[t] %[htbp] 
    \centering
    \vspace{-2mm}
    \includegraphics[width=0.96\linewidth]{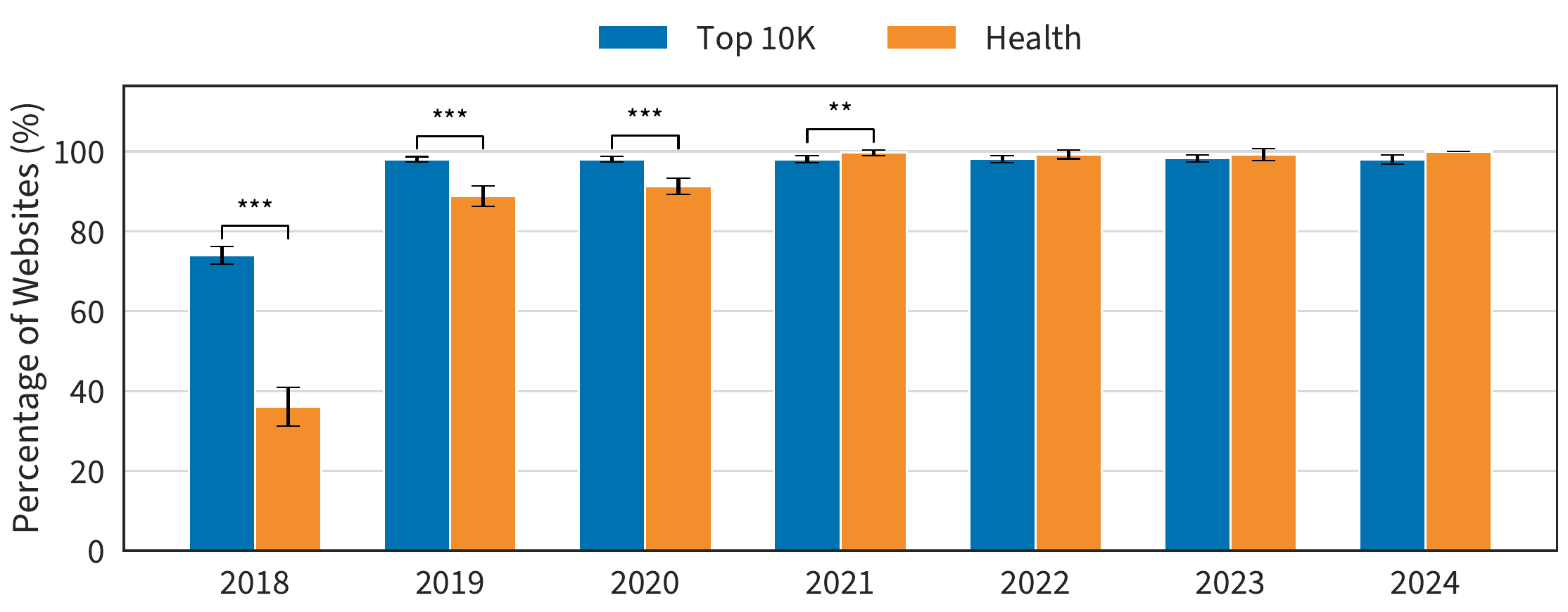} 
    \vspace{-4mm}
    \caption{Percentage of \textcolor{myblue}{control} and \textcolor{orange}{health} websites configuring \texttt{FirstPartyCookies} by year from 2018 to 2024. Error bars represent 95\% confidence intervals. Asterisks indicate statistical significance determined by a two-proportion z-test (**$p$ < 0.05, ***$p$ < 0.01). Effect sizes for significant results, measured using Cohen’s $h$, ranged from $h_{0.17}$ to $h_{0.78}$.} 
    \label{fig:arch-first-party-cookies} 
    \vspace{-2mm}
\end{figure}

Figure~\ref{fig:arch-first-party-cookies} shows that first-party cookie adoption surged between 2018 and 2019, soon after they were introduced in October 2018. 
%
% From 2021 through 2024, up to 98.4\% of all health and control websites configured first-party cookies.
%
From 2021 through 2024, over 98.0\% of control websites and over 99.2\% of health websites configured first-party cookies. 
Across both categories combined, the adoption reached upto 98.4\% in 2023.
This shows that advertisers largely adhere to the default setting of first-party cookies, rarely disabling them.

\noindent {\textbf{Automatic Advanced Matching (AAM).}}
AAM enables Meta to directly collect user attributes from websites.
Meta hashes user attributes before matching it to the corresponding attributes of all Facebook users \cite{Meta2025Advmatching}. 
Therefore, hashing user attributes does not limit Meta's ability to link the attributes to Facebook users \cite{FTC2024Hashing}. 
Moreover, hashing certain user attributes such as gender is meaningless since it has only two possible values (`M' or `F') as per Meta's documentation.

We conduct a longitudinal analysis of 11 user attributes that Meta collects as defined in the \texttt{selectedMatchKeys} array in the Pixel configuration script.
Figure~\ref{fig:arch-combined-aam} depicts the longitudinal trend of all 11 attributes.
Since the trend is largely similar, for the sake of brevity, we discuss two of these attributes here.
%
% Figure ~\ref{fig:arch-email-phone-AAM} illustrates a 
%
We observe a steady increase in Meta’s collection of user's email and phone attributes across control websites through 2024 and health websites up to 2021-2022.
%
% While AAM is not turned on by default, this increase can be attributed to Meta’s design of the AAM setup process, which leverages dark patterns to encourage its configuration.
%
While AAM is not enabled by default, this increase can be attributed to Meta’s AAM setup process, which exhibits design characteristics aligned with established dark pattern taxonomies \cite{gray2024darkpatternsCHI}.
During Pixel setup, Meta describes AAM as a means to ``enhance remarketing'' \cite{Meta2025Advmatching}.
Moreover, enabling AAM through a single toggle automatically activates the collection of all user information attributes by default, aligning with the \textit{Hiding Information} dark pattern as shown in Figure \ref{fig:arch-aam-toggle} \cite{gray2024darkpatternsCHI}.
To prevent specific identifiers from being collected, advertisers must expand the ``Show customer information parameters'' dropdown and manually opt-out for each user attribute.
Meta introduced country, date of birth, and external ID as new identifiers in 2020, further expanding the range of user attributes that it collects through AAM.

\begin{comment}
\begin{figure} [t] %[htbp] 
    \centering
    % \vspace{-2mm}
    \includegraphics[width=\linewidth]{figures/plot_Email_Phone_AAM_Stats.pdf} 
    \vspace{-7mm}
    \caption{Percentage of \textcolor{myblue}{control} and \textcolor{orange}{health} websites with Automatic Advanced Matching (AAM) configured for email and phone from 2018 to 2024. For trends related to all AAM keys, refer to Figure~\ref{fig:arch-combined-aam}. Asterisks indicate statistical significance determined by a two-proportion z-test (*$p$ < 0.1, **$p$ < 0.05, ***$p$ < 0.01). Effect sizes for significant results, measured using Cohen’s $h$, ranged from $h_{0.10}$ to $h_{0.44}$ for email and from $h_{0.11}$ to $h_{0.32}$ for phone.}
    %Note the general rise and the recent divergence, particularly the decrease on hospital sites for many keys.
    \label{fig:arch-email-phone-AAM} 
   \vspace{-4mm}
\end{figure}
\end{comment}

%%
The decline in AAM adoption among health websites beginning in 2023 can be linked to regulatory actions targeting tracking of potentially sensitive health information.
In 2022, HHS published the bulletin about the use of tracking pixels on health websites~\cite{hhs2024HIPAAtracking}.
This was followed by warning letters issued by FTC and HHS in 2023 \cite{FTC2023HeartTracking}. 
This and associated enforcement actions may have prompted health website advertisers to stop using AAM.
As of 2024, 47.8\% of health websites continue to still use AAM.

%%
% \vspace{-2mm}
\subsection{Tracking Restrictions}
\label{sec:archaeology-controls}

\noindent {\textbf{Unwanted Data.}} 
We detected that the \texttt{UnwantedData} configuration reflects filtering of URL and custom data parameters based on \texttt{blacklisted\_keys} (specified in plain text) and \texttt{sensitive\_keys} (specified as a SHA-256 hash). 
A key challenge in analyzing these filtering rules is the hashing of \texttt{sensitive\_keys}. 
We attempt to reverse SHA-256 hashed \texttt{sensitive\_keys} using a public database of pre-computed hashes available at CrackStation \cite{crackstation2025,Marie2017BitcoinSpringer}. 
This allowed us to successfully reverse 73.8\% of the \texttt{sensitive\_key} hashes. 
Overall, we find 4,136 unique \texttt{blacklisted} and 898 unique decrypted \texttt{sensitive} keys. 
%
% \vspace{-5mm}

\begin{figure}[htbp] 
    \centering
    \vspace{-2mm}
    \includegraphics[width=0.8\linewidth]{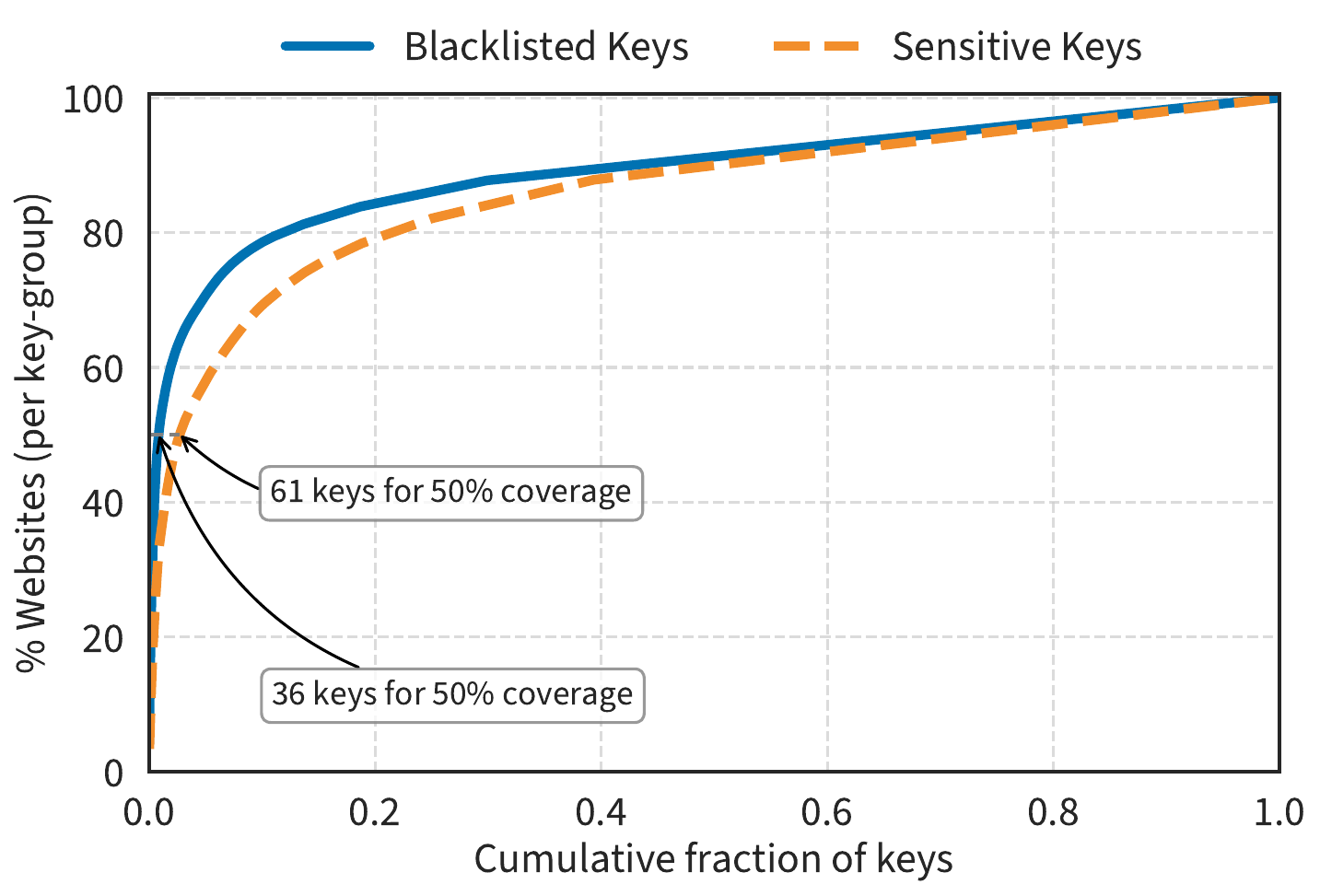} 
    \vspace{-3mm}
    \caption{Cumulative distribution of websites containing a given fraction of keys in \textcolor{myblue}{blacklisted} or \textcolor{orange}{sensitive} groups. A steep curve highlights that a small fraction of keys are present across a large proportion of websites, indicating concentrated key distribution.}
    \label{fig:arch-unwanted-overlap} 
    \vspace{-2mm}
\end{figure}

Our analysis revealed an overlap in \texttt{blacklisted\_keys} and \texttt{sensitive\_keys} across websites. 
As shown in Figure~\ref{fig:arch-unwanted-overlap}, half of the websites that contain \texttt{blacklisted} keys include at least one of the 36 common keys and half of the websites that contain \texttt{sensitive} keys include at least one of the 61 common keys.
This overlap in commonly observed keys suggests that Meta is likely responsible for selecting \texttt{blacklisted} and \texttt{sensitive} keys. 
This is consistent with Meta’s claim of detecting and filtering potentially prohibited information automatically~\cite{meta2025prohibitedinfoDocs}.

We also identify 143 keys (e.g., \texttt{searchTerm}, \texttt{locationName}, \texttt{q}) that are common across \texttt{blacklisted} and \texttt{sensitive} keys. 
However, despite similarities, there are clear differences between \texttt{black- listed} and \texttt{sensitive} keys. 
Blacklisted keys (specified in plaintext) frequently contained substrings explicitly indicative of common PIIs, such as \texttt{name}, \texttt{address}, \texttt{password}, \texttt{em} (email), \texttt{dob} (date of birth), \texttt{phone}, \texttt{IP}, \texttt{lat} (latitude), and \texttt{long} (longitude).
These PII-related substrings appear in 58.4\% of the unique \texttt{blacklisted} keys but only 7.3\% of the decrypted \texttt{sensitive} keys.
In contrast, the hashed \texttt{sensitive} keys include parameters related to potentially sensitive health information.
Examples include \texttt{doctor} (\textit{towerhealth.org}), \texttt{specialty} (\textit{balladhealth.org}), \texttt{height} (\textit{menningerclinic.org}), \texttt{gender} (\textit{doctor.webmd.com}), \texttt{hospital} (\textit{jeffersonhealth.org}), \texttt{lgbtq} (\textit{ucihealth.org}), \texttt{pregnant} (\textit{investing.com}) and \texttt{physician} (\textit{templehealth.org}).

\begin{figure}[htbp] % htbp
    \vspace{-3mm}
    \centering
    \includegraphics[width=0.96\linewidth]{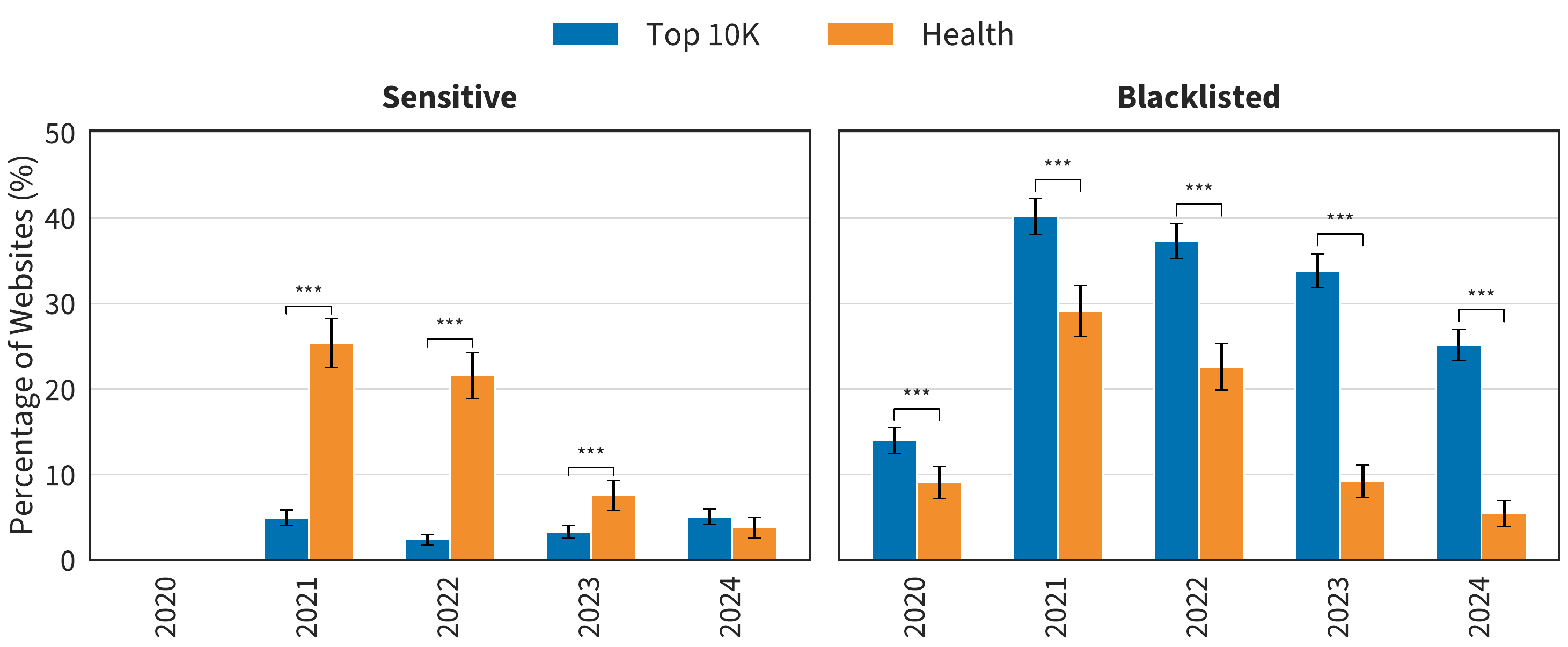}
    \vspace{-4mm}
    \caption{Percentage of control and health websites containing sensitive and blacklisted keys. Asterisks indicate statistical significance determined by a two-proportion z-test (***$p$ < 0.01). Effect sizes for significant results, measured using Cohen’s $h$, ranged from $h_{0.19}$ to $h_{0.66}$ for sensitive keys and from $h_{0.15}$ to $h_{0.63}$ for blacklisted keys.}
    \label{fig:sensitive_percentages}
    \vspace{-2mm}
\end{figure}

Figure \ref{fig:sensitive_percentages} shows that no \texttt{sensitive} keys were detected in Pixel configurations in 2020. 
From 2021 and 2022, health websites exhibited higher use of \texttt{sensitive} keys compared to control websites, with parameters such as \texttt{txtSearch}, \texttt{searchstr}, \texttt{childId}, \texttt{donor}, \texttt{gender}, \texttt{queryfilter}, etc. % \texttt{sortorder},
Over time, the gap between health and control websites in terms of use of \texttt{sensitive} keys narrows. 
One of the potential reasons could be heightened regulatory scrutiny on health websites. 
We next analyze the names of custom events observed in \texttt{black- listed} or \texttt{sensitive} keys as reflected in the \texttt{UnwantedData} configuration (see Listing 4 in Section \ref{sec:meta-pixel-configurations}). 
These custom event names suggest tracking of potentially sensitive health information such as ``OCD'' and ``PTSD Quiz'' (\textit{rogersbh.org}), ``Low Testosterone Form Submit'' (\textit{houstonmethodist.org}), ``CardiovascularSurgery'' (\textit{mountsinai.org}), and specific treatments like ``ImpressionGileadhiv'' (\textit{everydayhealth.com}).

\noindent {\textbf{Core Setup.}} 
We detected that the \texttt{protectedDataMode} configuration reflects Core Setup. 
Recall that Core Setup restricts the data collection by Meta Pixel, particularly for health, finance, and consumer reports websites deemed sensitive by Meta \cite{meta2025prohibitedinfoDocs}.
While Core Setup was officially introduced in May 2024, we first observe the \texttt{protectedDataMode} configuration in Pixel configurations in July 2023.
This suggests that Meta may have started deploying Core Setup on certain websites due to regulatory scrutiny even before its official announcements. 
This is consistent with advertisers reporting their pixels being placed under Core Setup \cite{Reddit2024Coresetup}.

\begin{figure}[htbp] %[htbp] 
    \centering
    \includegraphics[width=0.7\linewidth]{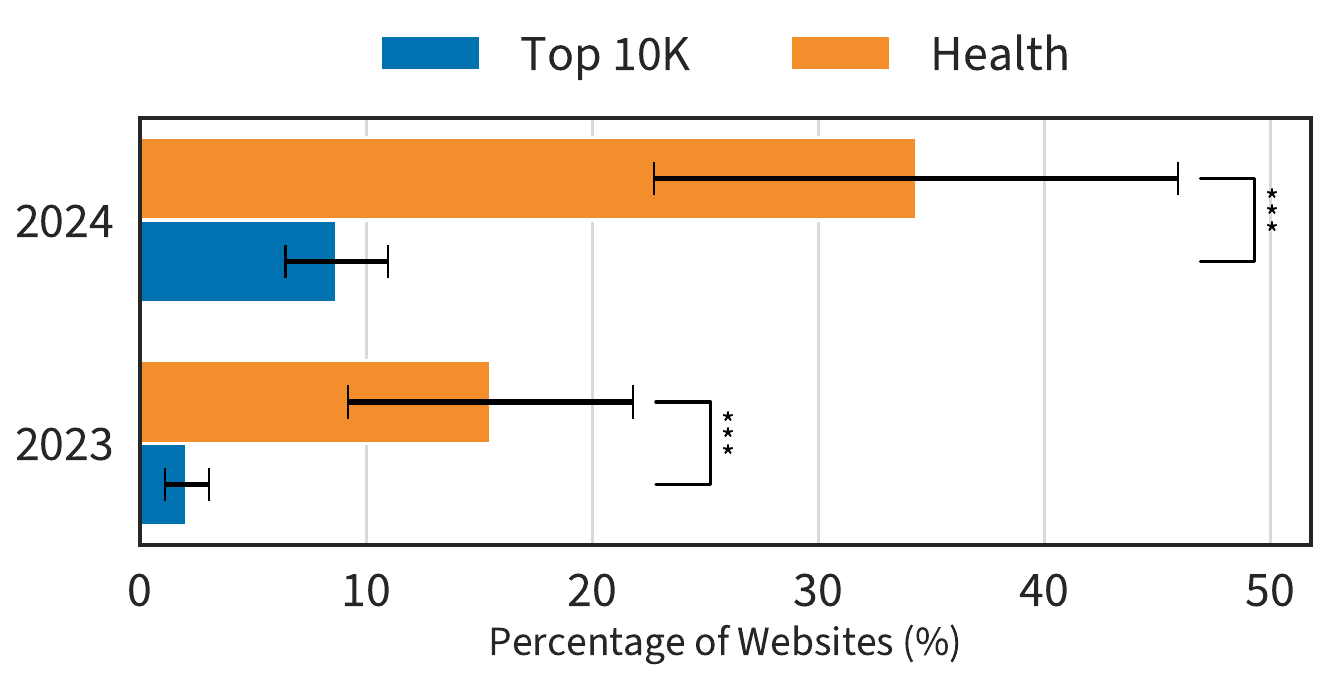} 
    \vspace{-3mm}
    \caption{Percentage of \textcolor{myblue}{control} and \textcolor{orange}{health} websites using \texttt{ProtectedDataMode} (i.e., Core Setup) over 2023 and 2024. Error bars represent 95\% confidence intervals. Asterisks indicate statistical significance determined by a two-proportion z-test (***$p$ < 0.01). Effect sizes for significant results, measured using Cohen’s $h$, ranged from $h_{0.52}$ to $h_{0.65}$.}
    \label{fig:arch-protected-data-mode} 
    \vspace{-1mm}
\end{figure}

Figure~\ref{fig:arch-protected-data-mode} shows a rise in Core Setup adoption from 2023 to 2024, with a notably steeper increase among health websites.
In 2023, only 2.1\% of control websites were under Core Setup, increasing to 8.7\% in 2024.
In contrast, Core Setup adoption rose sharply for health websites from 15.6\% in 2023 to 34.3\% in 2024.
While pixels on many health websites are in Core Setup, a majority have still not been placed under Core Setup.
For example, Meta Pixel continues to collect search terms on \textit{myamericannurse.com} in \texttt{s}, \textit{dshs.texas.gov} in \texttt{content}, and \textit{health.csuohio.edu} in \texttt{keys} parameters.
% \vspace{-3mm}

Once a pixel is placed under Core Setup, we observe that a JSON object listing the website's custom conversion and custom audience rules appears in its configuration.
Note that advertisers can specify rules based on visited URLs or custom parameters to track conversions or create custom audiences for targeted advertising~\cite{meta2025CustomConversion,meta2025CustomAudiences}.
Among websites in Core Setup, 73.5\% and 92.3\% have at least one custom conversion rule and one custom audience rule, respectively.
For instance, the pixel on \textit{healthline.com} tracks users visiting URLs containing ``/health/boils-on-buttocks'', ``/health/vaginal-pimples'', ``urge-incontinence'', and ``infertility'' using these custom rules. 
Similarly, the pixel on \textit{vitals.com} tracks users visiting URLs containing ``erectile-dysfunction'', ``cannabis'', and ``adhd'' through these custom rules.
% Similarly, the pixel on \textit{psychcentral.com} tracks users visiting URLs containing ``adhd'' and ``nicotine'' through these custom rules. 
%
Note that the pixel on this site was not placed in Core Setup until August 2024.
We also observe cases where websites deploy multiple pixels, but not all are placed in Core Setup.
For instance, the \textit{nationaljewish.org} hospital deployed four Meta pixels in late 2023, of which only one was placed in Core Setup. 
The pixel placed in the Core Setup was configured with custom conversion and audience rules to track specific health conditions by detecting presence of ``…/lung-cancer-screening/thank-you'', or ``…/cardiac-rehabilitation-program'' in the URL.
Even when Core Setup is enabled and URL is stripped to the domain name, we observe pixels on some websites circumventing Core Setup restrictions by including hashed versions of the full URL. 
For instance, the pixel on \textit{wexnermedical.osu.edu} is placed in Core Setup but the SHA-256 hash of the full URL is shared with Meta in the ud[dl] parameter. 
This effectively circumvents Core Setup's tracking restriction.

%%

% \vspace{-3mm}
\section{Conclusion}
\label{sec:discussion}

Meta Pixel has been available for more than a decade and is used on approximately a quarter of the websites today. 
Prior measurement studies were limited to studying the prevalence of tracking pixels, such as the Meta Pixel. 
In this paper, we present a deeper dive into the installation \textit{and} configuration of Meta Pixel. 
Using \texttt{PixelConfig}, our framework to reverse-engineer Meta Pixel configurations, we conduct a longitudinal analysis of Meta Pixel configuration on health and a control set of websites from 2017 to 2024. 
Our work sheds light on how Meta Pixel's activity tracking, identity tracking, and tracking restriction features have been adopted and configured. 
%
% Overall, we find that websites stick to Meta Pixel's out-of-the-box configurations, driven in part by the  defaults and dark patterns that nudge advertisers to not change them. 
Overall, we find that websites stick to Meta Pixel’s out-of-the-box configurations, driven in part by the defaults and design choices that nudge advertisers to not change them.
%
% For example, Meta Pixel was configured to automatically collect button click and page meta-data on up to 98.4\% of websites and first-party cookies on up to 98.4\% of websites. 
Meta Pixel was configured to automatically collect button clicks and page metadata on up to 98.4\% of websites, and first-party cookies on up to the same proportion of websites.
We find evidence that Meta Pixel tracks potentially sensitive information on health websites such as user interactions related to booking medical appointments or clicking buttons associated with specific medical conditions (e.g., erectile dysfunction). 
Although Meta later introduced tracking restriction controls, such as Core Setup and Unwanted Data, in response to regulatory scrutiny, we find that adoption of these controls---again largely driven by Meta rather than websites---is incomplete, can be ineffective, and can be circumvented.
Despite incomplete archival coverage of Meta Pixel configurations in the Wayback Machine, our robustness checks confirm the validity of our findings (see \Cref{sec:robustness-check}).

Our work contributes to the Internet measurement literature on tracking pixels by presenting a reverse-engineering framework and its application to analyze Meta Pixel configurations using the Wayback Machine. 
While we tailor and apply \texttt{PixelConfig} for Meta Pixel, future work can explore adapting this approach to study the configuration behaviors of other tracking pixels.
%
% the underlying methodology may be used to study the configurations of other tracking pixels.
%
To facilitate future research on tracking pixels, we have open-sourced \texttt{PixelConfig} and released the data 
%(list of health care provider websites and pixel source code) 
at \textit{\url{https://github.com/Yash-Vekaria/pixel-config}}.
% \textit{\url{https://anonymous.4open.science/r/pixel-config}}.

%%

\begin{acks}

This work was supported in part by the National Science Foundation
under award numbers 2138139 and 2103439.

\end{acks}

% \newpage
\vspace{-1mm}
\bibliographystyle{ACM-Reference-Format}
\bibliography{references}

\appendix

% \newpage
\section{Ethics}
% https://conferences.sigcomm.org/imc/2025/submission-instructions/}

%%
This study complies with ethical guidelines for research involving data collection and usage. 
Our research methodology is based on the analysis of publicly available data obtained from the Internet Archive’s Wayback Machine and controlled experiments conducted on a researcher-managed test website.
We observe Internet Archive's access limits to avoid overwhelming their infrastructure. 
The primary data sources we analyze include archived Meta Pixel configuration scripts and archived website snapshots, all of which are public records. 
Our research did not involve the collection, interception, or processing of any actual user's Personally Identifiable Information (PII). 
We strictly analyzed publicly accessible code and configuration data as deployed by websites. 
Thus we strongly believe that our work does not pose any ethical concerns.

\section{Top 10K Categorization}
\label{sec:website-categorization}

We categorized the top 10,000 websites using the WhoisXML Website Categorization API \cite{whoisxml}, which classifies domains based on their primary business or content category using a proprietary machine learning model trained on large-scale web data. Each website in our dataset was queried through the API, which returned one or more standardized category labels (e.g., Technology \& Computing, Healthcare Industry) if it could confidently categorize the website.
Table \ref{tab:website_categories_basic} shows that while 64.24\% of the websites were left Uncategorized, the majority of the remaining sites were classified under Business and Finance (32.36\%), Shopping (27.11\%), and Retail Industry (27.11\%), reflecting the commercial nature of much of the web traffic. The Health category, which was consolidated from multiple health-related subcategories (e.g., Healthcare Industry, Pharmaceutical Industry, Nutrition, Wellness, and Eldercare), accounted for only 2.04\% of the total websites.

\begin{table}[H]
\centering
\footnotesize
\vspace{-3mm}
\caption{Proportions of Website Categories in the Top 10K}
\vspace{-3mm}
\label{tab:website_categories_basic}
\renewcommand{\arraystretch}{1}
\begin{tabular}{l r}
\hline
\textbf{Category} & \textbf{Percentage (\%)} 
\\
\hline
Uncategorized                     & 64.24 \\
Business and Finance              & 32.36 \\
Shopping                          & 27.11 \\
Retail Industry                   & 27.11 \\
Style \& Fashion                  & 19.81 \\
Men's Clothing                    & 17.00 \\
Travel                            & 15.93 \\
Household Supplies                & 13.05 \\
Women's Clothing                  & 11.73 \\
Women's Accessories               & 10.06 \\
Home \& Garden                    & 9.83  \\
Personal Finance                  & 9.71  \\
Technology \& Computing           & 9.48  \\
Gifts and Greetings Cards         & 8.85  \\
Information Services Industry     & 8.01  \\
Technology Industry               & 7.85  \\
Telecommunications Industry       & 7.83  \\
Internet                          & 7.29  \\
Coupons and Discounts             & 6.95  \\
Business I.T.                     & 6.15  \\
Home Entertainment Systems        & 5.70  \\
Outdoor Decorating                & 5.68  \\
Remodeling \& Construction        & 5.36 \\
Home Entertaining                 & 5.29 \\
Family and Relationships          & 5.12 \\
Home Utilities                    & 5.11 \\
Business                          & 4.80 \\
Business Utilities                & 4.76 \\
Economy                           & 4.64 \\
Parenting                         & 4.30 \\
Budget Travel                     & 3.95 \\
Sensitive Topics                  & 3.80 \\
Careers                           & 3.29 \\
Honeymoons and Getaways           & 3.20 \\
Travel Preparation and Advice     & 2.86 \\
Entertainment Industry            & 2.55 \\
Events and Attractions            & 2.55 \\
Children's Clothing               & 2.52 \\
Personal Debt                     & 2.31 \\
Environmental Services Industry   & 2.31 \\
Personal Celebrations \& Life Events & 2.29 \\
Job Search                        & 2.27 \\
Hobbies \& Interests              & 2.18 \\
\textbf{Health}                   & \textbf{2.04} \\
\hline
\vspace{-2mm}
\end{tabular}
\end{table}

\vspace{-6mm}
\section{Error Bars}

For a given feature and in a given year, we analyze only those websites for which we find at least one configuration script in Wayback. 
Considering this subset, we compute the proportion p of websites that exhibit the feature in that year. 
We then construct a 95\% CI using a t-test based margin of error (plots are at \(p \times 100 \pm \mathrm{ME}\)):
\[
\mathrm{ME}
\;=\;
t_{0.975,\,n-1}
\;\sqrt{\frac{p\,(1-p)}{n}}
\;\times 100
\]
% Error bars in the plots are thus placed at \(p \times 100 \pm \mathrm{ME}\).

\noindent Note that similar to Lerner et al.~\cite{Lerner2016InternetjonesUSENIX}, we do not claim statistical significant comparisons but rather perform trend analysis of pixel configurations longitudinally on the web.

\vspace{-3mm}
\section{Other Pixel Configurations}

\begin{figure}[t]
    \centering
    \includegraphics[width=\linewidth]{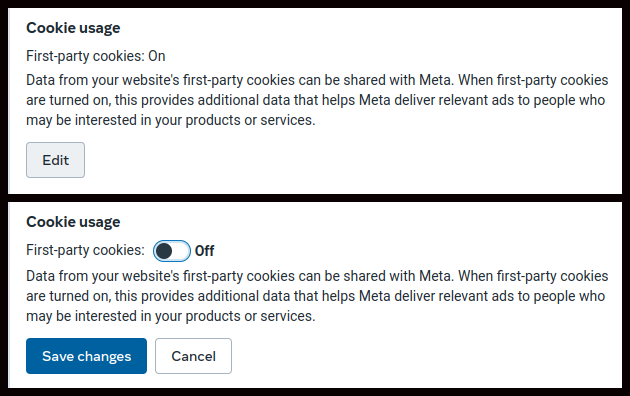}
    \caption{The complex process for enabling first-party cookies requires advertisers to first click `Edit', then toggle the setting, and finally press `Save Changes'.}
    \label{fig:arch-firstpartycookies-toggle}
    % \vspace{-15mm}
\end{figure}

\begin{figure}[htbp]  % H to force placement on the same page
    \centering
    \begin{minipage}{0.5\textwidth} % Adjust width to fit side by side
        \centering
        \includegraphics[width=\linewidth]{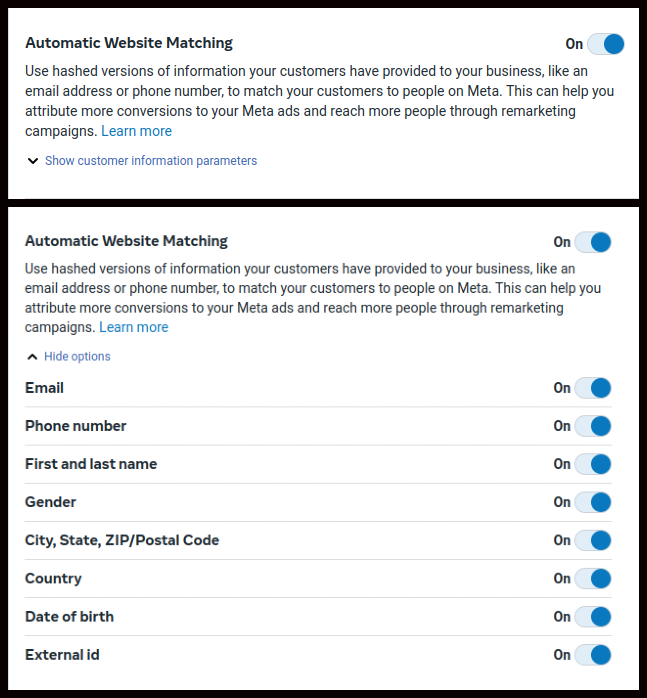}
        \caption{Activating the AAM toggle while the customer information parameters are concealed automatically enables all associated parameters.}
        \label{fig:arch-aam-toggle}
    \end{minipage}%
    \hfill
    \begin{minipage}{0.5\textwidth}  % Adjust width to fit side by side
        \centering
        \includegraphics[width=\linewidth]{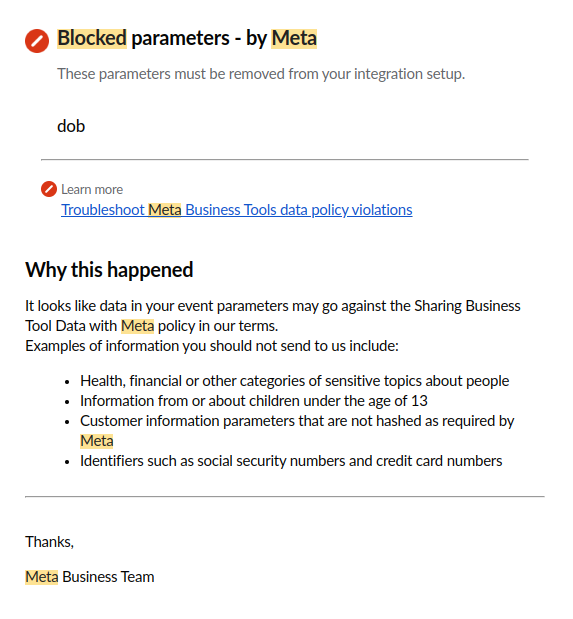}
        \vspace{-10mm}
        \caption{Email notification from Meta indicating that the `dob' parameter—configured on our controlled website to share users’ date-of-birth data—has been blocked.}
        \label{fig:arch-dob-blocked}
    \end{minipage}
\end{figure}

%%
% Figure~\ref{fig:arch-combined-aam} shows the longitudinal trend of all 11 Automatic Advanced Matching (AAM) keys corresponding to different user attributes.
%%
Figures \ref{fig:arch-firstpartycookies-toggle}, \ref{fig:arch-aam-toggle}, and \ref{fig:arch-dob-blocked} showcase UI controls in Meta Business Manager for first-party cookies, AAM, and parameter blocking by Meta. %, respectively.

\vspace{-1mm}
\section{Robustness}
\label{sec:robustness-check}

\begin{figure}[ht] % htbp
    \centering
    \includegraphics[width=0.95\columnwidth]{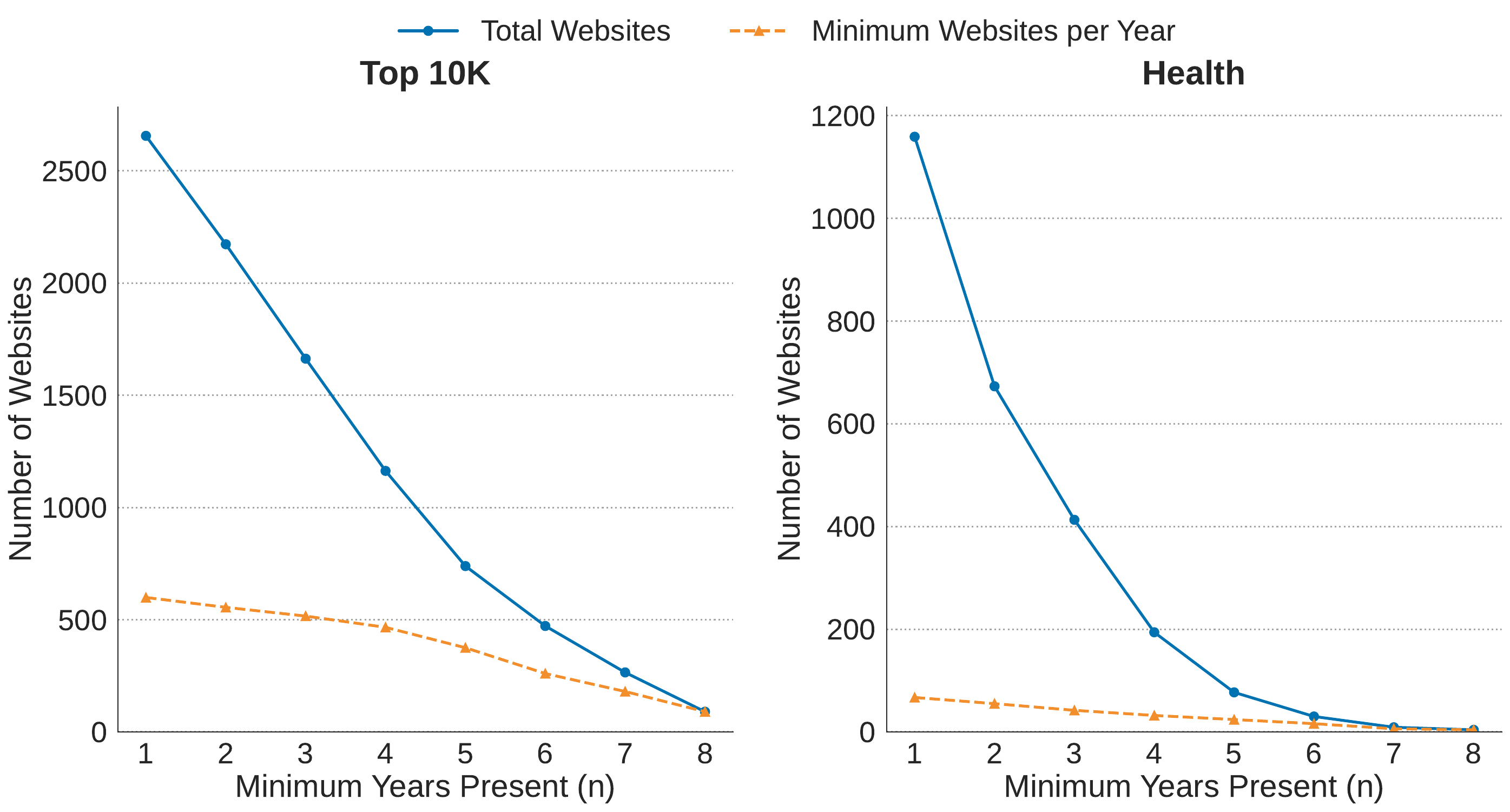} 
    \vspace{-3mm}
    \caption{Number of websites that appear in at least "n" years (blue), and the smallest number of websites available in any single year among those "n" years (orange), for the Top-10K (left) and Health (right) datasets.}
    \label{fig:stability-counts-robustness}
\end{figure}

Due to Wayback Machine's limitations in consistently archiving websites across the years, our longitudinal comparison did not comprise the same set of websites over the years to ensure enough sample size for trend analysis (Figure \ref{fig:stability-counts-robustness}). 
To assess whether missing configurations in any given year bias our results, we analyze the percentage adoption of different features using configurations fetched from live versions of websites (see Section \ref{subsec:validation}), where data from complete (Section \ref{subsec:live-robustness}).
Moreover, to validate whether variation in the set of websites across years affects our results, we also repeat the analysis using a stable cohort of websites present in at least four of the eight years, ensuring both temporal consistency and adequate sample size (Section \ref{subsec:stable-robustness}).

\vspace{-3mm}

\vspace{2mm}
\subsection{Live Websites}
\label{subsec:live-robustness}
Table \ref{tab:live_website_comparison} shows that the percentage adoption patterns among live top-10K ($n=101$) and health ($n=72$) websites remain largely consistent with our primary findings. First-party cookies are nearly universal across both groups ($\geq 98\%$). Tracking restrictions are substantially more common among health websites (Core Setup: $73.6\%$ vs.\ $6.9\%$), while automated events are less prevalent in health websites compared to top-10K sites ($27.8\%$ vs.\ $93.1\%$).

% \vspace{-5mm}
\begin{table}[ht]
\centering
\caption{Comparison of Meta Pixel Configurations on Live Health and Top-10k Websites (October 2025).}
\label{tab:live_website_comparison}
\footnotesize
\vspace{-2mm}
\renewcommand{\arraystretch}{1} % Adjusts row height for better readability
    \begin{tabular}{l r@{\,$\pm$\,}l r@{\,$\pm$\,}l l}
    \hline
    \textbf{Configuration} & \multicolumn{2}{c}{\textbf{Top-10k (\%)}} & \multicolumn{2}{c}{\textbf{Health (\%)}} & \textbf{$p < 0.01$} \\
    \hline
    \multicolumn{6}{l}{\textbf{Activity Tracking}} \\
    Inferred Events & 93.10 & 5.01 & 27.80 & 10.50 & Yes \\
    Event Setup Tool & 46.50 & 9.85 & 43.10 & 11.60 & No \\
    \hline
    \multicolumn{6}{l}{\textbf{Identity Tracking}} \\
    First-Party Cookies & 98.00 & 2.75 & 98.60 & 2.75 & No \\
    AAM-{em} (Email) & 16.67 & 8.76 & 68.32 & 9.18 & Yes \\
    AAM-{ph} (Phone) & 16.67 & 8.76 & 66.34 & 9.33 & Yes \\
    AAM-{fn} (First Name) & 15.28 & 8.45 & 63.37 & 9.51 & Yes \\
    AAM-{ln} (Last Name) & 15.28 & 8.45 & 63.37 & 9.51 & Yes \\
    AAM-{db} (Date of Birth) & 11.11 & 7.38 & 41.58 & 9.73 & Yes \\
    AAM-{ge} (Gender) & 13.89 & 8.13 & 62.38 & 9.56 & Yes \\
    AAM-{zp} (Zip Code) & 15.28 & 8.45 & 61.39 & 9.61 & Yes \\
    AAM-{ct} (City) & 15.28 & 8.45 & 61.39 & 9.61 & Yes \\
    AAM-{st} (State) & 15.28 & 8.45 & 61.39 & 9.61 & Yes \\
    AAM-{country} & 9.72 & 6.96 & 40.59 & 9.69 & Yes \\
    AAM-{external\_id} & 9.72 & 6.96 & 41.58 & 9.73 & Yes \\
    AAM-{fbp\_as\_external\_id} & 88.89 & 7.38 & 56.44 & 9.79 & Yes \\
    \hline
    \multicolumn{6}{l}{\textbf{Tracking Restrictions}} \\
    Unwanted Data (Sensitive) & 12.90 & 11.50 & 45.80 & 6.50 & Yes \\
    Unwanted Data (Blacklisted) & 83.20 & 7.30 & 26.40 & 10.20 & Yes \\
    Core Setup & 6.93 & 5.00 & 73.60 & 10.40 & Yes \\
    \hline
    \end{tabular}
\end{table}

% \vspace{-8mm}
\begin{table}[H]
\footnotesize
\centering
\caption{\#websites with Meta Pixel installations and corresponding configuration scripts found through the Wayback.}
\vspace{-2mm}
\label{tab:appendix-website-counts-basic}
\begin{tabular}{|c|c|c|c|c|}
\hline
\textbf{} & \multicolumn{2}{c|}{\textbf{Top 10k Websites}} & \multicolumn{2}{c|}{\textbf{Health Websites}} \\
\cline{2-5}
\textbf{Year} & \textbf{\makecell{Pixel\\Found}} & \textbf{\makecell{Configuration\\Found}} & \textbf{\makecell{Pixel\\Found}} & \textbf{\makecell{Configuration\\Found}} \\
\hline
2017 & 1,317 & 818 & 622 & 91 \\
2018 & 1,899 & 1,564 & 1,114 & 383 \\
2019 & 2,094 & 1,793 & 1,613 & 580 \\
2020 & 2,079 & 1,763 & 1,895 & 748 \\
2021 & 1,923 & 955 & 2,206 & 302 \\
2022 & 1,688 & 902 & 2,121 & 259 \\
2023 & 1,442 & 827 & 1,626 & 129 \\
2024 & 1,240 & 599 & 1,267 & 67 \\
\hline
\end{tabular}
\end{table}

%%
% \newpage
\subsection{Stable Websites}
\label{subsec:stable-robustness}
\begin{figure}[hbp] %[htbp] 
    \centering
    \includegraphics[width=0.9\linewidth]{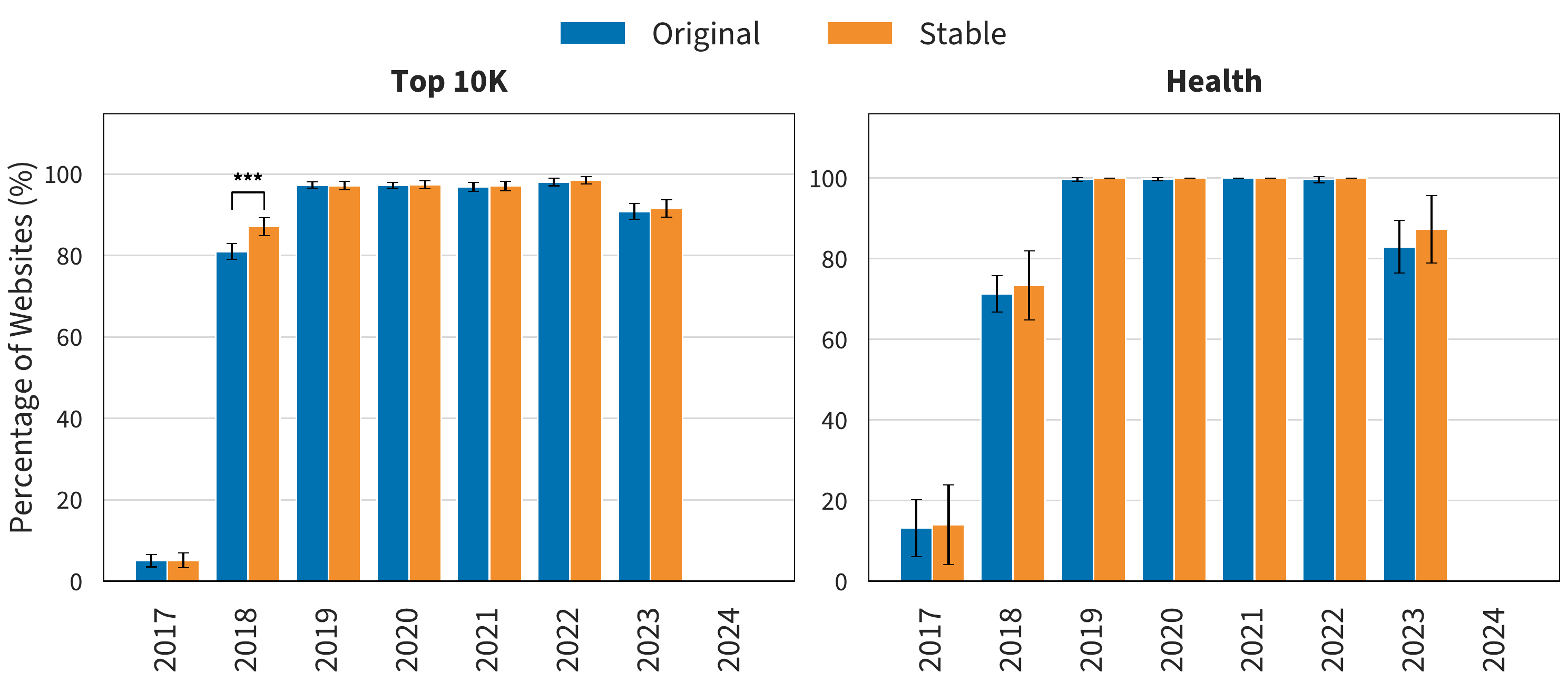} 
    \vspace{-4mm}
    \caption{A comparison of adoption for the \texttt{AutomaticSetup} configuration between the full dataset and a stable cohort of websites (with a snapshot for at least 4 out of 8 years) for Top 10K (left) and Health (right) websites. Error bars represent 95\% CI. Asterisks indicate statistical significance determined by a two-proportion z-test (*$p$ < 0.1, **$p$ < 0.05, ***$p$ < 0.01).}
    \label{fig:stability-robustness-check-automatic-setup} 
\end{figure}

\begin{figure} [htbp] %[htbp] 
    \centering
    \includegraphics[width=0.9\linewidth]{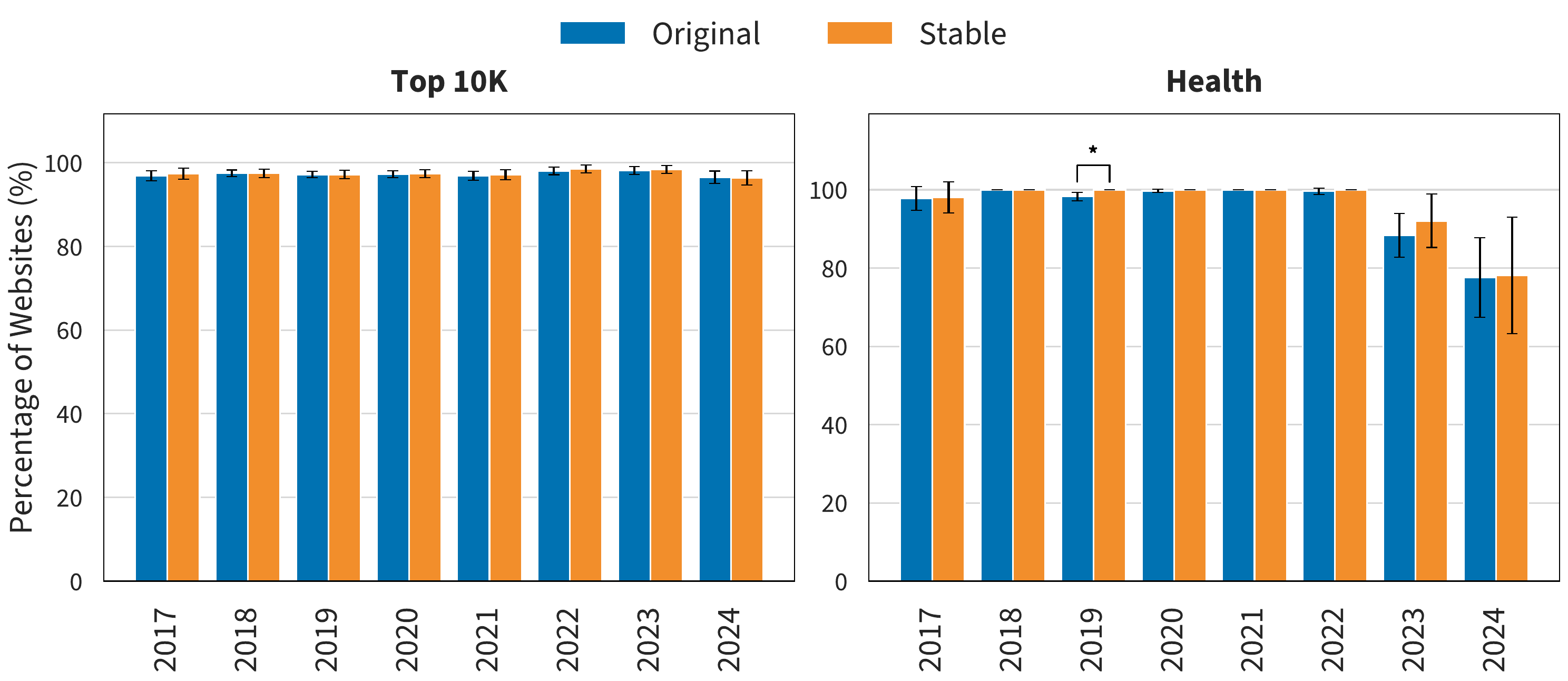} 
    \vspace{-3mm}
    \caption{A comparison of adoption for the \texttt{InferredEvents} configuration between the full dataset and a stable cohort of websites (with a snapshot for at least 4 out of 8 years) for Top 10K (left) and Health (right) websites. Error bars represent 95\% CI. Asterisks indicate statistical significance determined by a two-proportion z-test (*$p$ < 0.1, **$p$ < 0.05, ***$p$ < 0.01).}
    \label{fig:stability-robustness-check-inferred-events} 
\end{figure}

\begin{figure}[htbp]
    \centering
    \includegraphics[width=0.9\linewidth]{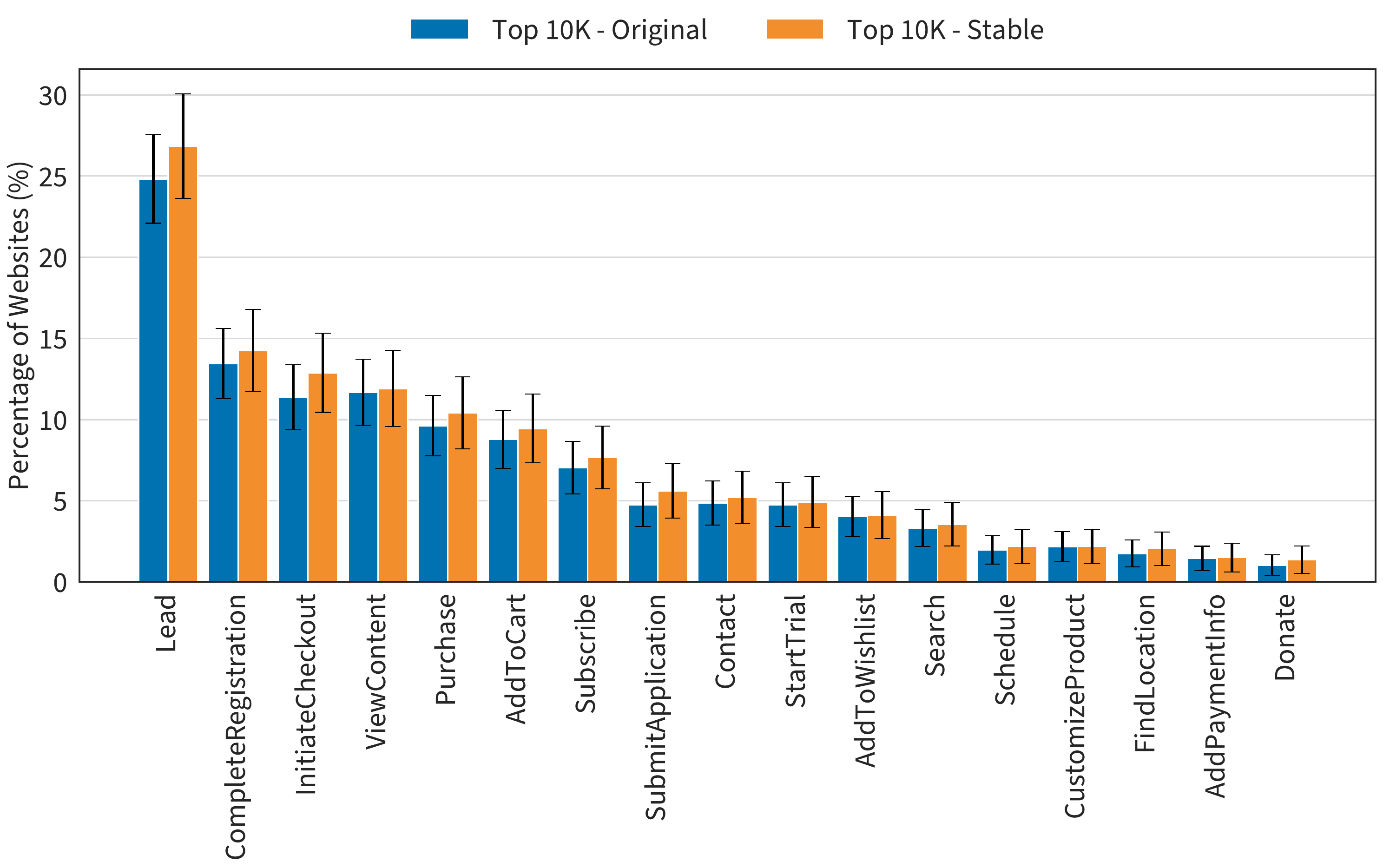}
    \vspace{-3mm}
    \caption{Adoption of standard events (Event Setup Tool) in the Top 10K dataset, comparing the full dataset to a stable cohort of websites (with a snapshot for at least 4 out of 8 years). Error bars represent 95\% confidence intervals. Asterisks indicate statistical significance determined by a two-proportion z-test (*$p$ < 0.1, **$p$ < 0.05, ***$p$ < 0.01).}
    \label{fig:stability-top10k-events}
\end{figure}

\begin{figure}[htbp]
    \centering
    \includegraphics[width=0.9\linewidth]{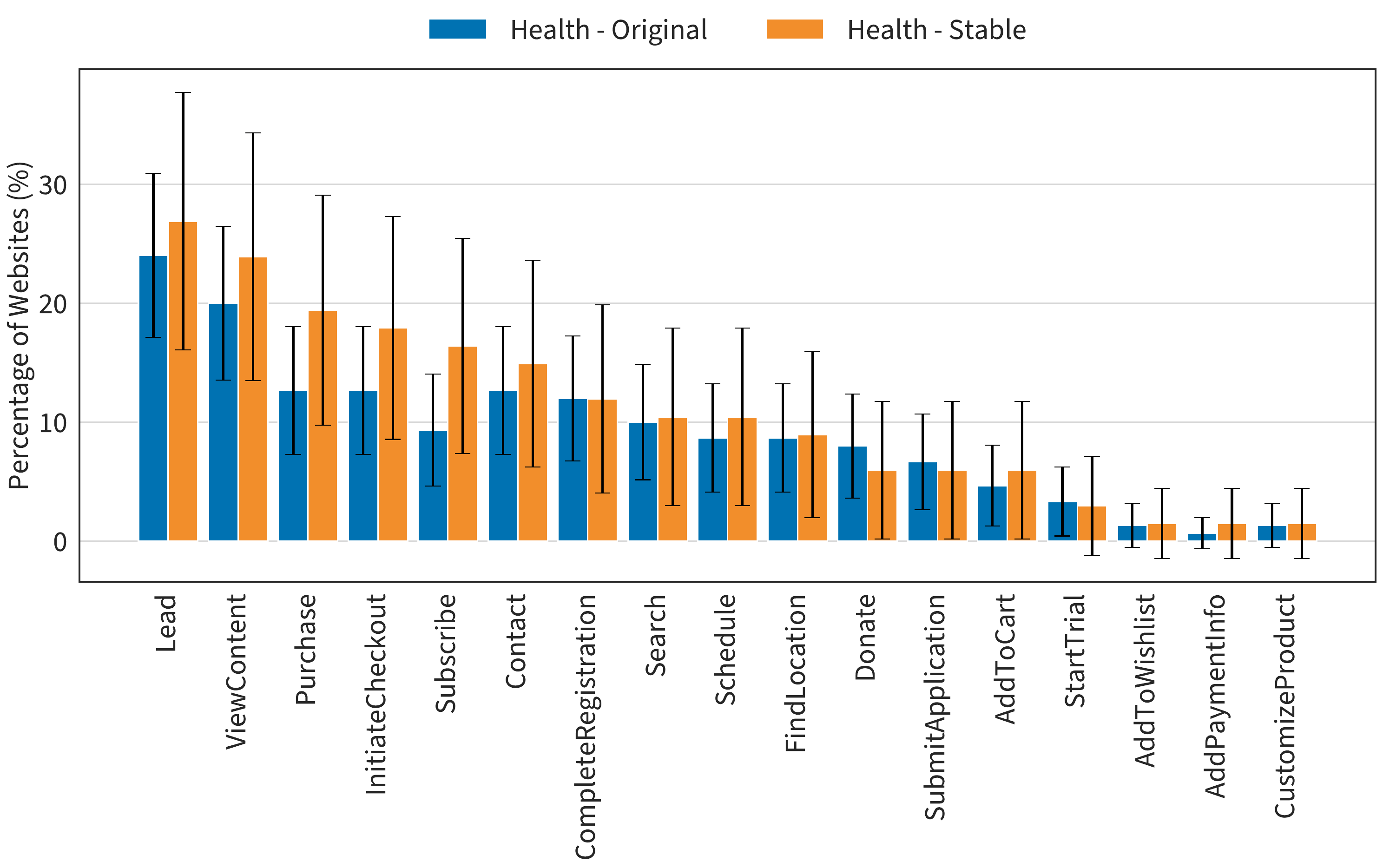}
    \vspace{-3mm}
    \caption{Adoption of standard events (Event Setup Tool) in Health websites, comparing the full dataset to a stable cohort of websites (with a snapshot for at least 4 out of 8 years). Error bars represent 95\% confidence intervals. Asterisks indicate statistical significance determined by a two-proportion z-test (*$p$ < 0.1, **$p$ < 0.05, ***$p$ < 0.01).}
    \label{fig:stability-health-events}
\end{figure}

\begin{figure*}[t] Unwanted Data

    \begin{subfigure}{0.49\textwidth}
        \includegraphics[width=\linewidth]{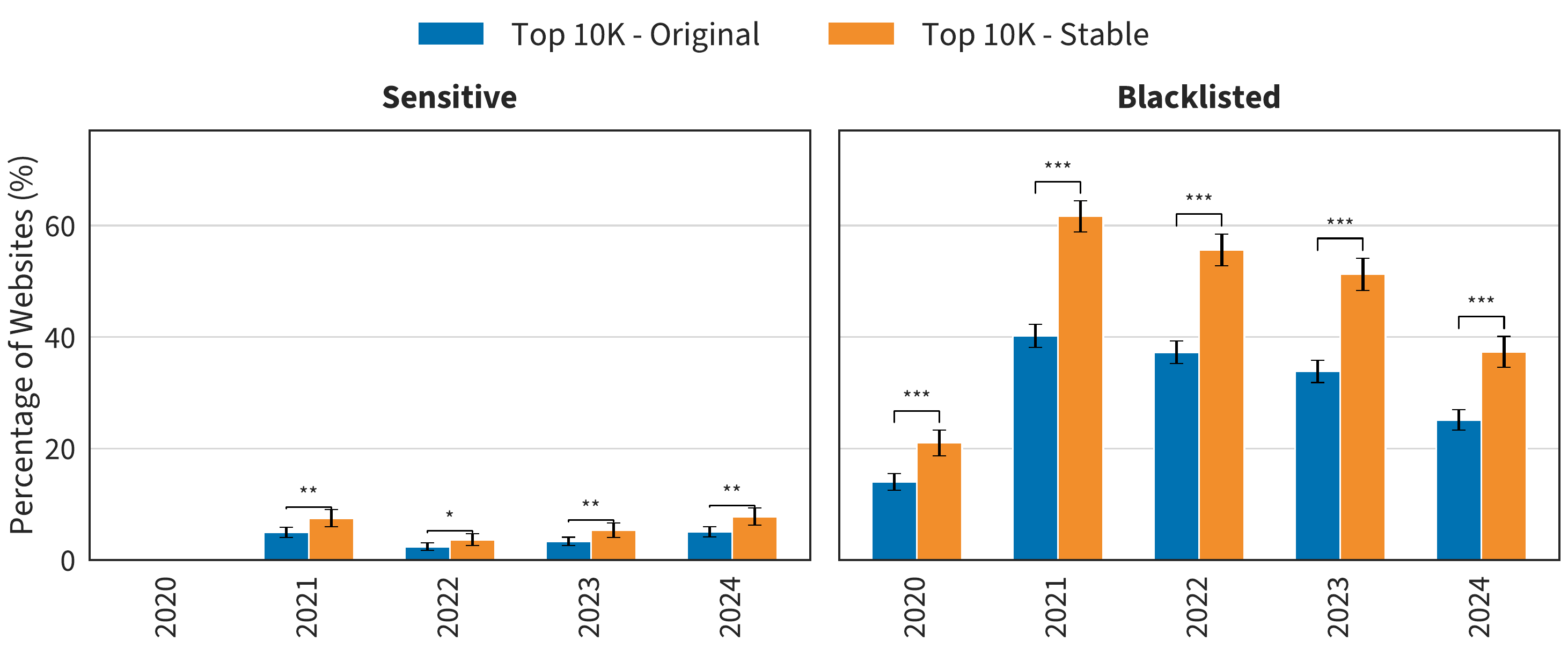}
        \caption{Top 10K} 
        \label{fig:stability-top10k-unwanted}
    \end{subfigure}
    \hfill 
    \begin{subfigure}{0.49\textwidth}
        \includegraphics[width=\linewidth]{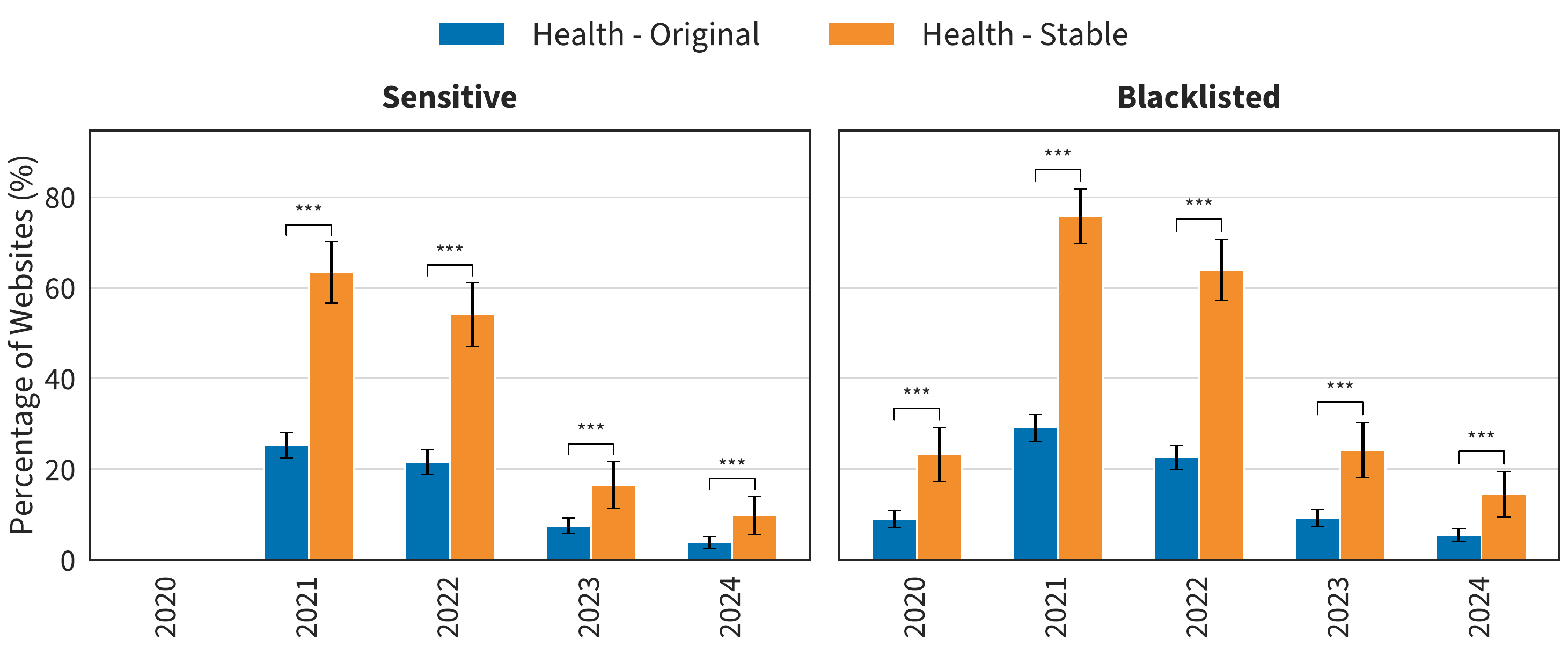}
        \caption{Health} 
        \label{fig:stability-health-unwanted}
    \end{subfigure}
    \caption{A comparison of adoption trends of sensitive and blacklisted keys between the full dataset and a stable cohort of websites present for at least 4 out of 8 years for Top 10K (left) and Health (right) websites. Error bars represent 95\% confidence intervals. Asterisks indicate statistical significance determined by a two-proportion z-test (*$p$ < 0.1, **$p$ < 0.05, ***$p$ < 0.01)}
    \label{fig:stability-robustness-check-unwanted}
\end{figure*}

Figures \ref{fig:stability-robustness-check-automatic-setup}–\ref{fig:stability-robustness-check-unwanted} compare configuration adoption in the stable cohort and the full dataset. Although adoption rates vary slightly, the year-to-year trends remain consistent.
%
% Figures \ref{fig:stability-robustness-check-automatic-setup} to \ref{fig:stability-robustness-check-unwanted} present the percentage adoption of configurations for the stable cohort of websites compared to the full dataset. While some variation in adoption rates is observed, the relative trends across years remain largely consistent.

\vspace{2.5mm}

\vspace{-2mm}

\vspace{-2mm}

\vspace{-2mm}

\begin{figure*}[hbp] Automatic Advanced Matching (Top 10K)
% [t] for top and * for full width
    \centering
    \begin{minipage}{\linewidth}  % Wrap both figures in a single minipage
        % First figure in the minipage (full width)
        \centering
        \includegraphics[width=\linewidth]{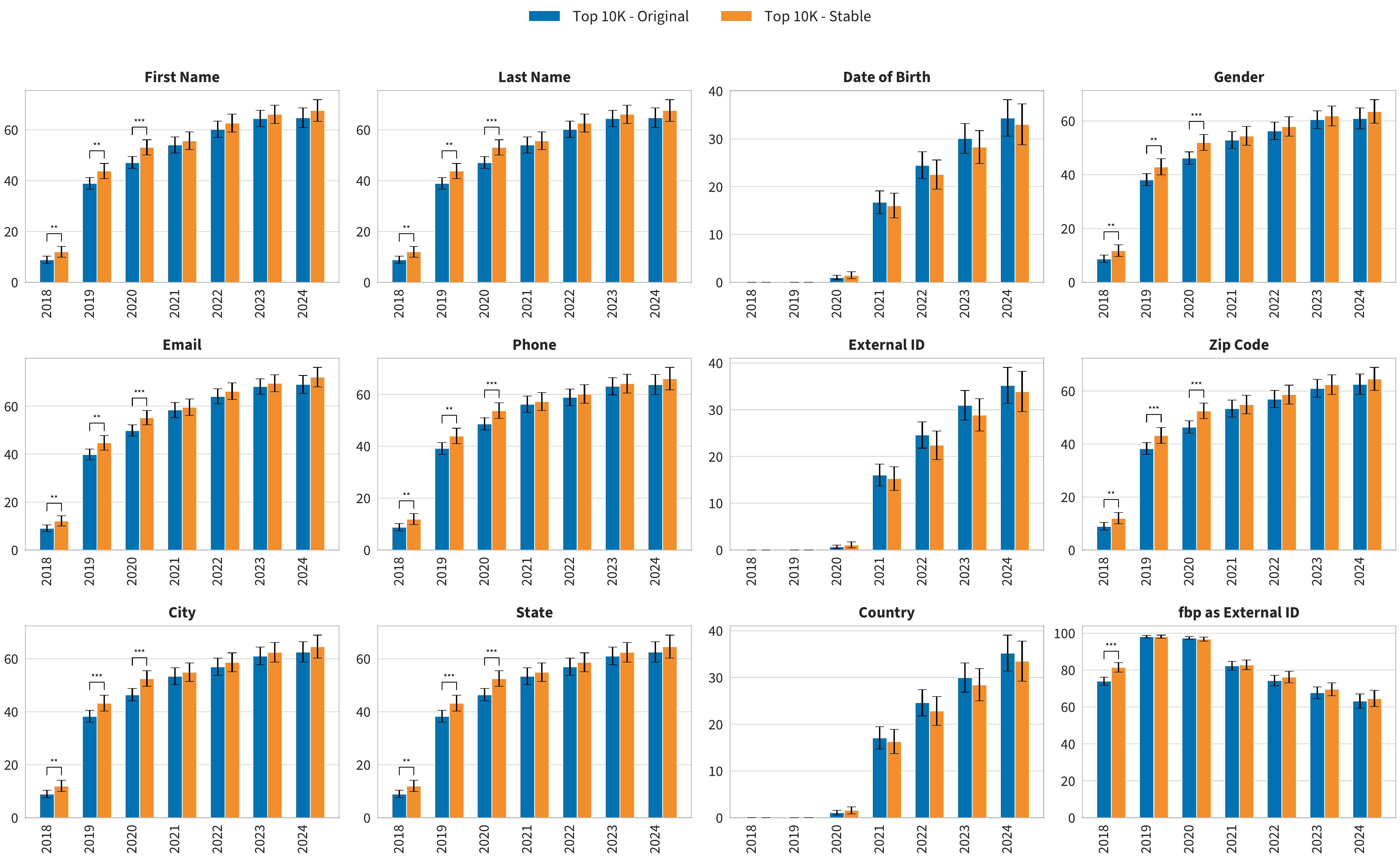}
        \caption{A comparison of adoption trends of specific match keys configured for the Automatic Advanced Matching (AAM) configuration between the full dataset and a stable cohort of websites present for at least 4 out of 8 years for Top 10K websites. Error bars represent 95\% confidence intervals. Asterisks indicate statistical significance determined by a two-proportion z-test (*$p$ < 0.1, **$p$ < 0.05, ***$p$ < 0.01)}
        \label{fig:arch-robustness-aam-top10k}
        \vspace{0.5cm}  % Adjust space between the two figures
    \end{minipage}
\end{figure*}

\begin{figure*}[htbp] Automatic Advanced Matching (Health)
% [t] for top and * for full width
    \centering
    \begin{minipage}{\linewidth}  
        % First figure in the minipage (full width)
        \centering
        \includegraphics[width=\linewidth]{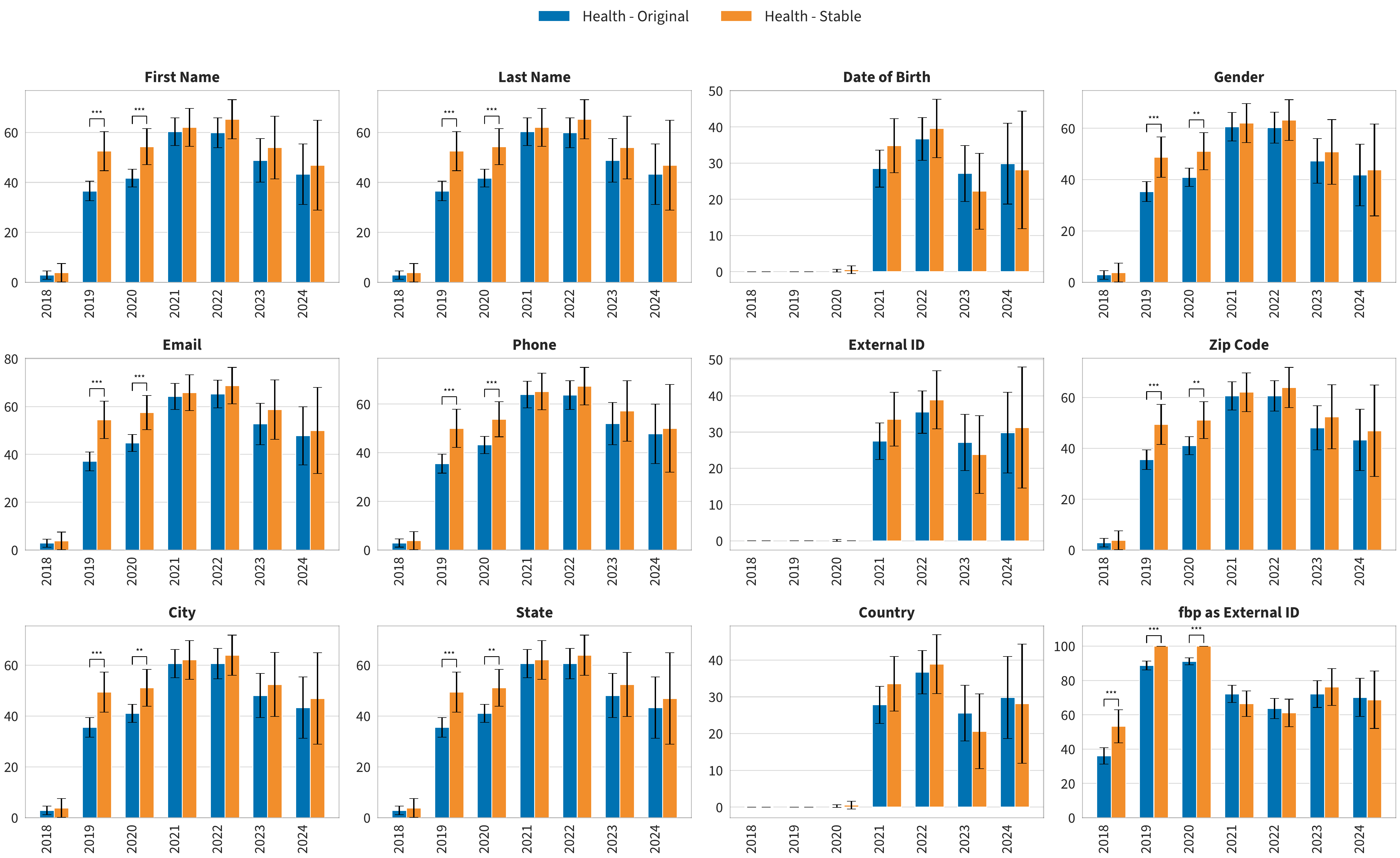}
        \caption{A comparison of adoption trends of specific match keys configured for the Automatic Advanced Matching (AAM) configuration between the full dataset and a stable cohort of websites present for at least 4 out of 8 years for Health websites. Error bars represent 95\% confidence intervals. Asterisks indicate statistical significance determined by a two-proportion z-test (*$p$ < 0.1, **$p$ < 0.05, ***$p$ < 0.01)}
        \label{fig:arch-robustness-aam-health}
        \vspace{0.5cm}  % Adjust space between the two figures
    \end{minipage}
\end{figure*}

\end{document}